# Frequency spectrum of biological noise:

# a probe of reaction dynamics in living cells

Sanggeun Song[1-3], Gil-Suk Yang[1-3], Seong Jun Park[1-3], Ji-Hyun Kim[1*], and Jaeyoung Sung[1-3*]

[1]Creative Research Initiative Center for Chemical Dynamics in Living Cells, Chung-Ang University, Seoul 06974, Korea.

[2]Department of Chemistry, Chung-Ang University, Seoul 06974, Korea.

[3]National Institute of Innovative Functional Imaging, Chung-Ang University, Seoul 06974, Korea.

**Author Information** The authors declare no competing financial interests. Correspondence and requests for materials should be addressed to J.S. (jaeyoung@cau.ac.kr) and J.-H.K. (jihyunkim@cau.ac.kr).

# Abstract

Even in the steady-state, the number of biomolecules in living cells fluctuates dynamically; and the frequency spectrum of this chemical fluctuation carries valuable information about the mechanism and the dynamics of the intracellular reactions creating these biomolecules. Although recent advances in single-cell experimental techniques enable the direct monitoring of the time-traces of the biological noise in each cell, the development of the theoretical tools needed to extract the information encoded in the stochastic dynamics of intracellular chemical fluctuation is still in its adolescence. Here, we present a simple and general equation that relates the power-spectrum of the product number fluctuation to the product lifetime and the reaction dynamics of the product creation process. By analyzing the time traces of the protein copy number using this theory, we can extract the power spectrum of the mRNA number, which cannot be directly measured by currently available experimental techniques. From the power spectrum of the mRNA number, we can further extract quantitative information about the transcriptional regulation dynamics. Our power spectrum analysis of gene expression noise is demonstrated for the gene network model of luciferase expression under the control of the *Bmal 1a* promoter in mouse fibroblast cells. Additionally, we investigate how the non-Poisson reaction dynamics and the cell-to-cell heterogeneity in transcription and translation affect the power-spectra of the mRNA and protein number.

**Introduction**

Fluctuation in the number of chemical species is ubiquitous and particularly important in small reactors such as living cells. This chemical fluctuation persists even in the steady-state, and its stochastic dynamics carries valuable information about the mechanism and the dynamics of intracellular reaction processes. The development of single molecule fluorescence imaging techniques has made it possible to directly monitor the time-traces of the protein number in each individual cell [1-3]. A great deal of research has focused on the *magnitude* of protein or mRNA number variation among genetically identical cells [4]. On the other hand, the *dynamics* of the protein or mRNA population fluctuation has been rarely the subject of investigation.

Taking Gillespie's chemical Langevin equation (CLE) approach [5], Simpson, Cox, and Saylor theoretically investigated the effects of feedback regulation on the protein number power spectrum, or the Fourier transform of the time correlation function (TCF) of the protein number fluctuation [6-8]. Bratsun, Volfson, Tsimring, and Hasty then investigated the power spectrum of the protein noise when the protein decay rate or the feedback regulated gene expression rate at a given time is dependent on the protein density at an earlier time, using the chemical master equation (CME) [9]. Tănase-Nicola, Warren, and Wolde presented a relationship between the power spectra of the input and output signals for various biochemical networks [10], starting from the CLE of biochemical networks. The CLE provides an efficient description of the stochastic chemical dynamics by separating the reaction rate into two parts, one systematic and the other noisy. However, how the analytical results from the CLE can be derived from the CME or its generalization has not been explicitly shown yet. In addition, it is still not well-understood how the mechanism and the dynamics of a product creation process are related to stochastic properties in the product creation rate and the TCF or the power spectrum of the product number fluctuation.

It was recently shown that intracellular chemical fluctuation follows a simple mathematical rule, called the Chemical Fluctuation Theorem [4,11], in its generation and propagation through reaction networks in living cells. According to the Chemical Fluctuation Theorem, the time-correlation function (TCF) of the mRNA copy number or the translation rate is the determining factor of the magnitude of the non-Poisson protein noise. The TCF of the mRNA copy number is, in turn, dependent on the mechanism and dynamics of the transcription process, during which RNA polymerases synthesize mRNAs. However, thus far, the TCF of the mRNA number fluctuation in living cells has yet to be reported in the literature, due to a lack of the proper experimental tools needed to obtain the time traces of the mRNA copy number in each cell.

According to Onsager's regression hypothesis [12], the TCF of the mRNA copy number must have the same relaxation dynamics as the mean mRNA number near the equilibrium or steady-state, so that the TCF of the mRNA copy number carries the same information as the relaxation dynamics of the mean mRNA number. Onsager's hypothesis is indeed correct for the conventional birth-death network model [13]. However, it is questionable whether his hypothesis also holds for *vibrant* reaction networks in living cells, where the rate coefficients can be stochastic variables that differ from cell to cell and fluctuate over time due to their coupling to the cell state. Regardless of whether or not Onsager's hypothesis is correct for intracellular reaction networks, we have yet to understand how, in general, the TCF or the power spectrum of intracellular chemical fluctuation is related to the topology and the dynamics of intracellular reaction networks.

In this work, we present a simple equation for the product number power spectrum, considering a general model of reaction processes in living cells, which enables us to convert the power spectrum of the product number (PSPN) to the power spectrum of the product creation rate (PSPR) in the case where product decay is a unimolecular process with a constant rate. Using this result, we investigate the PSPN and the PSPR for three different models of birth-death processes, and show that the PSPR, obtained from the PSPN, is a sensitive probe of the mechanism and dynamics for the product creation process. We find that the PSPR is a monotonically decaying function of frequency for a multi-channel process and a non-monotonic function with one or more peaks for a multi-step reaction process, but vanishes for a single channel Poisson process.

By applying this theory to the gene expression network model, we obtain the power spectra and the TCFs of mRNA number and the transcription rate from the time traces of the protein number in each cell. This is significant because the time traces, or the power spectra, of the mRNA number or the transcription rate are difficult to obtain by currently available experimental techniques. From the power spectrum of the mRNA number, we can extract quantitative information about the gene regulating promoter dynamics. We demonstrate our power spectrum analysis of gene expression noise for the gene expression network model utilized by Naef and co-workers [14] to investigate the time traces of luciferase expressed under the control of *Bmal 1a* in mouse fibroblast cells. We also investigate the power spectra of the mRNA and protein number for a more general gene expression network model in which active gene transcription and translation are not simple rate processes. Throughout our investigation, we confirm the correctness of our theory against accurate stochastic simulation results.

**Theory**

The PSPN, $S_z(\omega)$, is defined as

$$S_z(\omega) \equiv \lim_{T \to \infty} T^{-1} \left\langle \left| \int_{-T/2}^{T/2} dt\, e^{-i\omega t} \delta z(t) \right|^2 \right\rangle, \tag{1}$$

where $\langle \cdots \rangle$ and $\delta z(t)$ denote the average over the number of the time trajectories of the product number and the deviation of the product number from the mean product number at time *t*, respectively. According to the Wiener-Khinchin theorem[15,16], the power spectrum defined in Eq. (1) is identically equal to the Fourier transform of the steady-state TCF of the product number fluctuation, i.e.,

$$S_z(\omega) = \int_{-\infty}^{\infty} dt e^{-i\omega t} \langle \delta z(t) \delta z(0) \rangle_{ss} . \qquad (2)$$

The functional form of the PSPN is dependent on the TCF of the product creation rate, which is, in turn, dependent on the mechanism and dynamics of the product creation process. We find that, for any birth-death network, the PSPN is given by

$$S_z(\omega) = \frac{2\langle R \rangle}{\omega^2 + \gamma^2} + \frac{S_R(\omega)}{\omega^2 + \gamma^2}, \qquad (3)$$

where $\langle R \rangle$, $\gamma$, and $S_R(\omega)$ respectively denote the mean product creation rate, the inverse lifetime of the product molecule, and the PSPR, that is, $S_R(\omega) = \int_{-\infty}^{\infty} dt e^{-i\omega t} \langle \delta R(t) \delta R(0) \rangle_{ss}$. See Supplemental Material 1 for the derivation of Eq. (3). This equation is consistent with the previous results obtained by using the CLE [7,8]. The first term on the right-hand side (R.H.S.) of Eq. (3) has the same functional form, regardless of the detailed mechanism and dynamics of the product creation process; this term is determined by only two parameters: the mean product creation rate and the inverse lifetime of the product molecule. In contrast, the functional form of the second term can vary, because $S_R(\omega)$, or the Fourier transform of the TCF of the product creation rate fluctuation, is dependent on the mechanism of the product creation network and the dynamics of the individual reaction processes that compose the network.

The PSPR, $S_R(\omega)$, can be easily extracted from the PSPN, $S_z(\omega)$. From Eq. (3), one obtains

$$S_R(\omega) = \left(\omega^2 + \gamma^2\right)\left[S_z(\omega) - S_z^0(\omega)\right] = 2\langle R\rangle\left[S_z(\omega)/S_z^0(\omega) - 1\right], \quad (4)$$

where $S_z^0(\omega)$ denotes the first term on the R.H.S. of Eq. (3). Given that the inverse lifetime, $\gamma$, of product molecules can be estimated independently, we can easily calculate the mean product creation rate, $\langle R\rangle$, from the mean product number, $\langle z\rangle$, by $\langle R\rangle = \langle z\rangle/\gamma$. With $\langle R\rangle$ and $\gamma$ at hand, we can convert the PSPN, $S_z(\omega)$, to the PSPR, $S_R(\omega)$, by Eq. (4) and $S_z^0(\omega) \equiv 2\langle R\rangle/\left(\omega^2 + \gamma^2\right)$.

The PSPR, $S_R(\omega)$, can be calculated for any reaction network model (see Supplemental Material). For example, in Fig. 1, we show $S_R(\omega)$ and the corresponding TCF of the product creation rate for three different product creation processes: the simple one-step process, the multi-channel process, and the multi-step process. In the simplest case, where the product creation reaction is a simple Poisson process, the product creation rate is constant in time, so that $\langle \delta R(t)\delta R(0)\rangle = S_R(\omega) = 0$. When the product creation process is a multi-channel reaction process, shown in Fig. 1b, the TCF of the product creation rate is given by a multi-exponential function of time, and the corresponding power spectrum is a monotonically decaying function of frequency, $\omega$ (see Eqs. (M3-5) and (M3-6) in Supplemental Material). In contrast, for a multi-step reaction process, shown in Fig. 1c, the TCF of the product creation rate becomes an oscillatory function of time as the step number increases, so that the power spectrum is a non-monotonic function with one or potentially more peaks (see Eq. (M3-12)). The derivation of the exact analytic expressions of the PSPR, $S_R(\omega)$, are presented for both the multi-channel reaction process and the multi-step reaction process in Supplementary Method 3.

The correctness of our theoretical results is confirmed against stochastic simulation results, as shown in Fig. 1d-g. The PSPN data, $S_z(\omega)$, calculated from the simulated product number time traces by Eq. (1) are converted to the PSPR data, $S_R(\omega)$, by Eq. (4). As shown in Fig. 1, these $S_R(\omega)$ data obtained from the simulated product number time traces are in perfect agreement with the prediction of our analytic results for all three reaction models in Fig. 1a-c. Both the theoretical prediction and the simulation results show that the power spectrum, $S_R(\omega)$, of the product creation rate is a more sensitive probe of the mechanism and dynamics of the product creation process than the PSPN, $S_z(\omega)$. This is because $S_z(\omega)$ is always contributed from $S_z^0(\omega)$, which is independent of the microscopic details of the product creation process, whereas $S_R(\omega)$ is not contributed from $S_z^0(\omega)$.

In the special case where the product creation process is a renewal process [17,18], in which the dynamics of one reaction event is not affected by the previous reaction history, we find that the PSPR is related to the distribution, $\psi(t)$, of the intermittent times between successive product creation events by

$$S_R(\omega) = 2\langle R \rangle \mathrm{Re}\left[\frac{\tilde{\psi}(\omega)}{1-\tilde{\psi}(\omega)}\right], \tag{5}$$

where $\tilde{\psi}(\omega)$ designates $\int_0^\infty dt e^{-i\omega t}\psi(t)$. For a single enzyme reaction network composed of elementary renewal processes, the analytic expression of $\tilde{\psi}(\omega)$ can be obtained from the method presented in ref [17]. This equation enables us to calculate the power spectrum for a single enzyme renewal process. It is known that the reaction process of multiple enzymes is not a renewal process[19,20]. However, one can show that the PSPR for a system of *n* enzymes is simply given by $nS_R(\omega)$, given that the correlation between different enzymes is negligible.

Equations (3) and (4) are applicable to the case where the product creation process is not a renewal process as well. A representative example of a non-renewal process is the gene expression process; rates of the chemical processes constituting gene expression can be a stochastic variable coupled to various cell state variables, such as promoter strength, the population of gene machinery proteins and transcription factors, the phase in the cell cycle, and the nutrition state, forcing the stochastic property of gene expression to deviate from a simple renewal process[4,21].

**Application to gene expression**

We demonstrate an application of Eqs. (3) and (4) to a quantitative analysis of the time traces of the protein copy number using the gene expression network model shown in Fig. 2a and b. This model has been used by Naef and coworkers to investigate the stochastic time traces of the number of luciferase expressed under the *Bmal 1a* promoter in mouse fibroblast cells [14]. According to the authors, under the *Bmal 1a* promoter, the transcription process involves a gene activation process composed of 7 intermediate reaction steps, which has highly sub-Poisson characteristics, and a simple one-step gene deactivation process with Poisson characteristics [14,22]. The one-step gene deactivation and multi-step gene activation model is schematically represented in Fig. 2b. For this gene expression network model, we conduct a stochastic simulation to obtain the time traces of the protein copy number, and then use the simulation data to calculate the protein number power spectrum. We compare this protein number power spectrum with the theoretical prediction of Eq. (3) for the same gene expression network model. As shown in Fig.3c, the prediction of Eq. (3) is in perfect agreement with the simulation results. An explicit, analytic result for the protein number power spectrum obtained from Eq. (3) is presented in Supplementary Information 5.

By applying Eq. (4) to the translation process, one can obtain the mRNA number power spectrum from the protein number power spectrum, despite the mRNA number power spectrum being difficult to directly obtain from currently available experimental tools. For the translation process, the protein creation rate, or the translation rate, is given by $R_{TL} = k_{TL} m$, where $k_{TL}$ and $m$ are the translation rate coefficient and the mRNA copy number, respectively. In general, $k_{TL}$ represents the translation rate per mRNA, and it is a stochastic variable dependent on various cell state variables, which include ribosome concentration, concentrations of amino acids, mRNA conformation, and so on. Let us first consider the simplest case where the fluctuation in $k_{TL}$ has negligible influence on the protein number fluctuation compared to the mRNA number fluctuation, which is found to be true in bacterial gene expression [1]. By applying Eq. (4) to the translation process, we obtain

$$k_{TL}^2 S_m(\omega) = (\omega^2 + \gamma_p^2)\left[S_p(\omega) - S_p^0(\omega)\right] = 2\langle R_{TL}\rangle\left[S_p(\omega)/S_p^0(\omega) - 1\right], \qquad (6)$$

where $S_p^0(\omega) = 2\langle R_{TL}\rangle/(\omega^2 + \gamma_p^2)$. As mentioned above, $\langle R_{TL}\rangle (= k_{TL}\langle m\rangle)$ can be easily estimated from the mean protein level and the inverse lifetime of protein, that is, $\langle R_{TL}\rangle = \langle p\rangle\gamma_p$. Given that the inverse lifetime, $\gamma_m$, of mRNA can be estimated separately, the value of $k_{TL}$ can also be estimated from the following asymptotic relation, $S_p(\omega)/S_p^0(\omega) - 1 \cong k_{TL}\gamma_m\omega^{-2}$, valid in the high frequency regime (see Supplementary Information 5). Even in the presence of the strong cell-to-cell variation in the translation rate coefficient, the relation between the mRNA number power spectrum and protein number power spectrum is similar to Eq. (6), as shown later in this work.

We use Eq. (6) to estimate the mRNA number power spectrum, $S_m(\omega)$, from the protein number power spectrum, $S_p(\omega)$, obtained from the stochastic simulation results. Alternatively, we can also directly calculate the mRNA number power spectrum by applying Eq. (3) to the transcription network or by stochastic simulation of the transcription network. As shown in Fig. 2e, all three results are in excellent agreement.

Applying Eq. (4) in a similar manner to the transcription process, we can convert the mRNA number power spectrum to the power spectrum of the transcription rate, i.e.,

$$S_{R_{TX}}(\omega) = (\omega^2 + \gamma_m^2)[S_m(\omega) - S_m^0(\omega)] = 2\langle R_{TX}\rangle[S_m(\omega)/S_m^0(\omega) - 1], \quad (7)$$

where $S_m^0(\omega)$ is defined as $S_m^0(\omega) = 2\langle R_{TX}\rangle/(\omega^2 + \gamma_m^2)$ with $\langle R_{TX}\rangle = \langle m\rangle\gamma_m$. As shown in Fig. 2f, the power spectrum of the transcription rate fluctuation extracted from the mRNA number power spectrum is independent of the mRNA lifetime, as it must be, although the mRNA number power spectrum is dependent on the mRNA lifetime. As can be seen in Fig. 2a and b, the power spectrum of the transcription rate obtained from the simulation results is in good agreement with the prediction of our analytic result obtained for the gene expression network model.

From the power spectrum of the transcription rate, one can further extract valuable information about the dynamics of the gene regulation process. For the transcription network model shown in Fig. 2a, the transcription rate can be represented by $R_{TX} = \xi k_{TX}$. Here, $\xi$ denotes the stochastic variable representing the promoter regulating gene state, whose value is 1 for the gene in the active state but 0 for the gene in the inactive state, and $k_{TX}$ denotes the active gene transcription rate. The power spectrum of the transcription rate is related to the lifetime distribution, $\psi_{on(off)}(t)$, of the active (inactive) gene state by

$$S_{R_{TX}}(\omega) = k_{TX}^2 S_\xi(\omega) = \frac{2k_{TX}^2}{\tau_{on} + \tau_{off}} \mathrm{Re} \frac{[1 - \hat{\psi}_{on}(s)][1 - \hat{\psi}_{off}(s)]}{s^2[1 - \hat{\psi}_{on}(s)\hat{\psi}_{off}(s)]}\bigg|_{s=i\omega}. \quad (8)$$

Here, $\hat{\psi}_{on(off)}(s)$ and $\hat{\psi}_{on(off)}(s)$ denote the power spectrum of the gene state variable, $\xi$, and the Laplace transform of the lifetime distribution of the active (inactive) gene state, respectively.

$\tau_{on(off)}$ designates the mean lifetime of the active (inactive) gene state, that is, $\tau_{on(off)} = \int_0^\infty dt\, t\psi_{on(off)}(t)$. We also obtain the analytic expression of $S_{R_{TX}}(\omega)$ for a more complex transcription model, in which the transcription process following gene activation is an arbitrary non-Poisson stochastic process, and present it in Supplemental Material.

The non-monotonic power spectrum of the transcription rate, shown in Fig. 2f, emerges because the gene activation process is a multi-step consecutive reaction process. For the gene activation-deactivation model shown in Fig. 2b, the deactivation of the active gene is a Poisson process, and $\psi_{on}(t)$ is a simple exponential function; in contrast, the gene activation process is a multi-step process with $\psi_{off}(t)$ being a non-monotonic, unimodal distribution (see Supplemental Material). For this model, $\hat{\psi}_{on}(s)$ and $\hat{\psi}_{off}(s)$ are given by $k_a/(s+k_a)$ and $\prod_{i=1}^{N} k_i/(s+k_i) \left[ \equiv \hat{f}(\mathbf{k},s) \right]$, respectively. According to ref [14], the activation process of *Bmal 1a* in the embryonic stem cells of mouse is best represented by 7 consecutive Poisson reaction processes, and the corresponding lifetime distribution, $\psi_{off}(t)$, of the inactive gene state involves 7 different rate parameters, the optimized values of which are given by $k_1 = k_2 = \cdots = k_6 = 9.93 \times 10^{-2}\, \text{min}^{-1}$ and $k_7 = 0.23\, \text{min}^{-1}$.

We find that the gamma distribution, $t^{a-1}e^{-t/b}/b^a \Gamma(a) \left[ \equiv f'(a,b,t) \right]$, serves as a good approximation of the non-Poisson lifetime distribution, $f(\mathbf{k},t)$, with the optimized parameter values. When $\psi_{off}(t)$ is a gamma distribution, $\hat{\psi}_{off}(s)$ is given by $(sb+1)^{-a}$. Respectively substituting $(sb+1)^{-a}$ and $k_a/(s+k_a)$ into $\hat{\psi}_{off}(s)$ and $\hat{\psi}_{on}(s)$ in Eq. (8), we obtain a new analytic result of $S_{R_{TX}}(\omega)$. We find that this result provides a good quantitative explanation of the power-spectra of the transcription rate shown in Fig. 2f, which are obtained from a simulation of the transcription network with $\hat{\psi}_{off}(s) = \hat{f}(\mathbf{k},s)$. $\psi_{on}(t)$ and $\psi_{off}(t)$ are optimized from the quantitative analysis of $S_{R_{TX}}(\omega)$ with use of $k'_a e^{-k'_a t}$ and $\psi_{off}(t) = f'(a,b,t)$, and as can be seen in Fig. 3a and b, the optimized $\psi_{on}(t)$ and $\psi_{off}(t)$ are in good agreement with the original model with $\psi_{on}(t) = k_a e^{-k_a t}$ and $\psi_{off}(t) = f(\mathbf{k},t)$.

It is worth mentioning that, when the rates of the $n$ consecutive reaction processes composing the gene activation process are the same, i.e., when $k_1 = k_2 = \cdots = k_n = k$, the gamma distribution, $f'(n,k^{-1},t)$, is the exact lifetime distribution, $\psi_{off}(t)[= f(\mathbf{k},t)]$, of the inactive gene state. For the transcription network model shown in Fig. 2a, the values of the seven rate parameters in the multi-step gene activation process do not greatly differ from each other. Therefore, the success of our analysis, using the gamma distribution as an approximate representation of $\psi_{off}(t)$, is not surprising.

The protein number power spectrum, which can be converted to the power spectrum of the mRNA with use of Eq. (6), is a more sensitive probe of transcription regulation than the steady-state protein or mRNA number distribution. As shown in Fig. 3c, the power spectrum of the mRNA number is quite sensitive to the reaction dynamics of the gene activation or deactivation process. For example, the mRNA number power spectrum shown in Fig. 3b defies explanation by the simple transcription network model, in which both gene activation and deactivation are simple Poisson processes. For this simple one-step Poisson gene activation model, the mRNA number power spectrum is a monotonically decaying function of frequency. However, the mRNA number power spectrum becomes a non-monotonic function with one or more peaks for the multi-step non-Poisson gene activation model, and its shape is strongly dependent on the number of intermediate steps composing the gene activation process, as shown in Fig. 3c. On the other hand, as shown in Fig. 3d, the gene expression network model with the simple one-step Poisson gene activation process yields the same steady-state protein number distributions as the gene expression network with the multi-step gene activation shown in Fig. 2a and b. This means that the power-spectrum of the transcription rate is a more sensitive probe of the mechanism and dynamics of the transcription process than the steady-state protein number distribution. The integration of the power spectrum, $S_q(\omega)$, of quantity $q$, over the entire frequency range is the same as the variance in quantity. In other words, the detailed shape of the power spectrum of mRNA or protein copy number carries more information than the variance in the mRNA or protein copy number, which is a commonly used experimental measure of gene expression noise.

Although active gene transcription and translation of each mRNA are often assumed to be simple Poisson processes in gene expression models in the literature, which is also assumed in Figs. 2a and 3a, these processes can be complicated, non-Poisson processes, because they are composed of multi-step or multi-channel elementary processes and because their reaction rates couple to heterogeneous cell environments [4]. With this in mind, from this point forward, we consider the mRNA and protein power spectra of the more general transcription network model shown in Fig. 4a, where active gene transcription is a non-Poisson process with rate being a dynamic stochastic variable, which differs from cell to cell and fluctuates over time.

The mRNA number power spectrum of the transcription network shown in Fig. 4a is dependent not only on the gene regulating promoter dynamics but also on the active gene transcription dynamics. For this model, the transcription rate can be represented by $R_{TX} = \xi k_{TX}(\Gamma)$. Here, $\xi$ and $k_{TX}(\Gamma)$ denote the gene state variable defined above Eq. (8) and the active gene transcription rate that is dependent on the cell state, $\Gamma$. The power spectrum, $S_{R_{TX}}(\omega)$, of the transcription rate for the model shown in Fig. 4a is given by

$$\tilde{S}_{R_{TX}}(\omega) = \tilde{S}_{\xi}(\omega) + \tilde{S}_{k_{TX}}(\omega) + \tilde{S}_{\xi}(\omega) * \tilde{S}_{k_{TX}}(\omega), \tag{9}$$

where $\tilde{S}_q(\omega)$ denotes $S_q(\omega)/\langle q \rangle^2$ (see Supplemental Material). In Eq. (9), $\tilde{S}_{\xi}(\omega)$ is related to $\tilde{S}_{\xi}(\omega) = \dfrac{2(\tau_{on} + \tau_{off})}{\tau_{on}^2} \text{Re} \left. \dfrac{[1-\hat{\psi}_{on}(s)][1-\hat{\psi}_{off}(s)]}{s^2[1-\hat{\psi}_{on}(s)\hat{\psi}_{off}(s)]} \right|_{s=i\omega}$ given in Eq. (8) by

$\tilde{S}_{\xi}(\omega) = (\tau_{on} + \tau_{off})^2 S_{\xi}(\omega)/\tau_{on}^2$, which has a non-monotonic frequency dependence when gene activation is a multi-step sub-Poisson process, as shown in Fig. 2f. The shape of $\tilde{S}_{k_{TX}}(\omega)$ is dependent on the active gene transcription dynamics, as is demonstrated in Fig. 1g. In Fig. 4, we present the power spectrum, $\tilde{S}_{R_{TX}}(\omega)\left[= 2\langle R_{TX}\rangle^{-1}(S_m(\omega)/S_m^0(\omega) - 1)\right]$, of the transcription rate when active gene transcription is a super-Poisson process such as the two-channel reaction process shown in Fig. 1b. For this model, $\tilde{S}_{k_{TX}}(\omega)$ is a monotonically decaying function of frequency (see Fig. 1g), given by $\tilde{S}_{k_{TX}}(\omega) = \lambda(\omega^2 + \lambda^2)^{-1}\eta_{k_{TX}}^2$ with $\lambda$ and $\eta_{k_{TL}}^2$ denoting the speed and relative variance in the active gene transcription rate fluctuation. As can be seen in Fig. 4b, when $\eta_{k_{TL}}^2 = 0.1$, the power spectrum, $\tilde{S}_{R_{TX}}(\omega)$, of the transcription rate is a non-monotonic function of frequency, nearly identical to Fig. 2f, and shows no strong dependency on the fluctuation speed, $\lambda$, in the active transcription rate. This is because, when $\eta_{k_{TL}}^2 \ll 1$, $\tilde{S}_\xi(\omega)$ originating from gene regulation of the promoter is the major contributor to $\tilde{S}_{R_{TX}}(\omega)$, while $\tilde{S}_{k_{TX}}(\omega)$ originating from active gene transcription is only a minor contributor (see Fig. 4c). However, as $\eta_{k_{TX}}^2$ increases, so too does the contribution from the active gene transcription dynamics to $S_{R_{TX}}(\omega)$, causing $S_{R_{TX}}(\omega)$ to significantly deviate from $S_\xi(\omega)$; $S_{R_{TX}}(\omega)$ is a monotonically decaying function of frequency in the low frequency regime due to the contribution from $S_{k_{TX}}(\omega)$, as demonstrated in Fig. 4b and c. When $\eta_{k_{TX}}^2$ is far greater than unity, $\tilde{S}_{R_{TX}}(\omega)$ is dominantly contributed from $\tilde{S}_{k_{TX}}(\omega)$ and $\tilde{S}_\xi(\omega) * \tilde{S}_{k_{TX}}(\omega)$, the last two terms on the R.H.S. of Eq. (9), and it is due to the latter contribution that $\tilde{S}_{R_{TX}}(\omega)$ has a non-monotonic frequency dependency in contrast to $\tilde{S}_{k_{TX}}(\omega)$.

In general, the translation rate coefficient, $k_{TL}$, is a random variable with an arbitrary cell-to-cell distribution. For this model as well, Eq. (3) exactly holds, and the protein number power spectrum is related to the mRNA number power spectrum by

$$S_p(\omega)/S_p^0(\omega) - 1 = \langle k_{TL} \rangle (1 + \eta_{k_{TL}}^2) S_m(\omega) / 2\langle m \rangle, \tag{10}$$

given that the cell-to-cell heterogeneity in $k_{TL}$ is much greater than the dynamic fluctuation of $k_{TL}$ in each cell (see Supplemental Material).

Naef and co-workers investigated the gene regulation dynamics of the *Bmal 1a* promoter in mouse fibroblast cells without performing differentiation. For this system, the gene copy number is always unity and does not vary. However, in a general gene expression system, gene copy number varies with time in a time scale much longer than the individual transcription event. In the simplest case where the time correlation between the number of proteins created by one gene and the number of proteins created by another is negligible, the power spectrum of the protein number is given by the R.H.S. of Eq. (7) multiplied by the average gene copy number. We leave it to future research to investigate how the cell-environment induced correlation between the expression levels of different genes affects the power-spectrum of mRNA and protein.

There are cases where the product decay process is not a simple Poisson process. For such cases, Eq. (3) and the equations that are derived from Eq. (3) in this work are only approximately valid. A generalization of this work to encompass the case where product decay is a non-Poisson process is possible, and we leave this for future research as well.

**Conclusion**

We investigate how the product number power spectrum is related to the topology of the product creation network and the dynamics of elementary processes composing the network. Our theory enables one to obtain the power spectrum of the product creation rate (PSPR) from the power spectrum of the product number (PSPN). We observe that the shape of PSPR is more sensitive to the mechanism and dynamics of the product creation process than the shape of PSPN. When the product creation process is a Poisson process, the PSPR vanishes; when the product creation process is a super-Poisson process, the PSPR is a monotonically decaying function of frequency; but when the product creation rate is a sub-Poisson process, the PSPR is a non-monotonic function of frequency with one or more peaks. The protein number power spectrum is a far more sensitive probe of the gene regulation dynamics than the steady-state protein number distribution or other common measures, such as the Fano factor and the relative variance of protein number variability. By applying our result to a gene expression network, we can extract the mRNA number power spectrum from the protein number power spectrum. From the mRNA number power spectrum, we can extract quantitative information about the gene regulation dynamics of the promoter and the active gene transcription dynamics. This is demonstrated for the gene network model of *luciferase* expression under the *Bmal 1a* promoter in mouse fibroblast cells.

# Acknowledgements

This research was supported by the Creative Research Initiative Project program (NRF-015R1A3A2066497) funded by the National Research Foundation (NRF) of the Korean government; the NRF grant (MSIP) (2015R1A2A1A15055664); and the Priority Research Center Program through the NRF (2009-0093817), funded by the Korean government.

# Figures

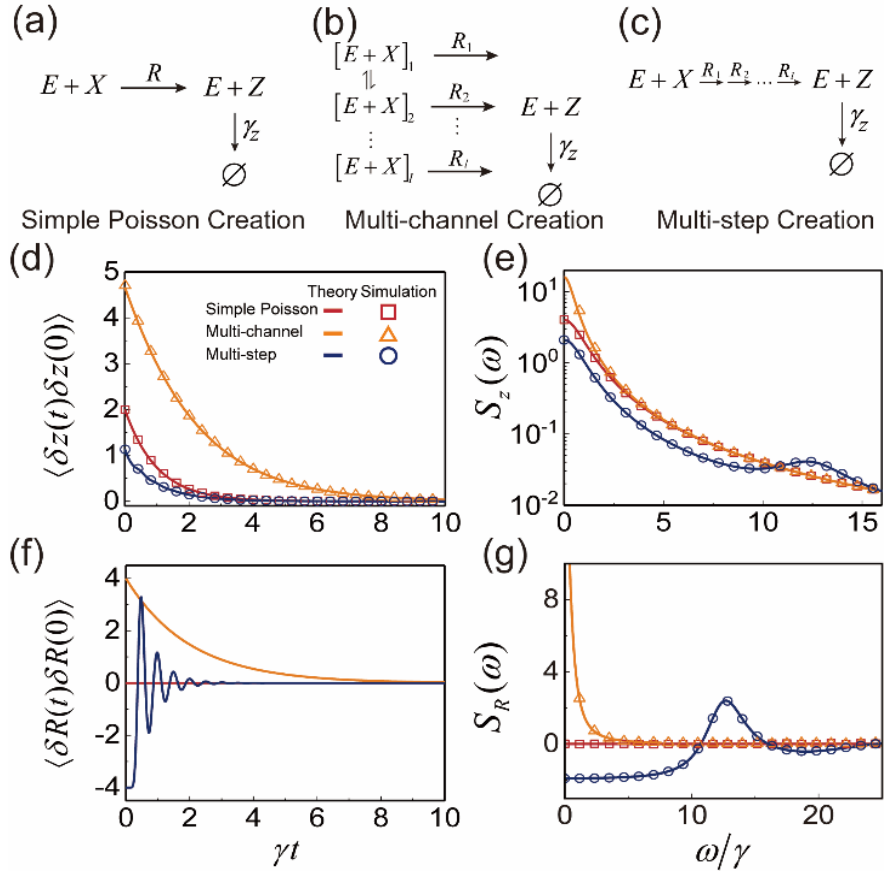

**Fig. 1. Time autocorrelation functions and power spectra of the product number and the product creation rate for three product creation-annihilation networks.** (a-c) Reaction schemes for the three different types of product creation processes: the simple Poisson process, the multi-channel process, and the multi-step process. The product decay process is a simple Poisson process. (d-e) Time autocorrelation function of the product number fluctuation and the power spectrum of the product number (PSPN) for the reaction schemes (a-c). (f-g) Time autocorrelation function of the product creation rate fluctuation and the power spectrum of the product creation rate (PSPR). The PSPR is related to the PSPN by Eq. (4). The PSPR is far more sensitive to changes in the reaction scheme than the PSPN. The PSPR is zero at all frequencies for Scheme a, a monotonically decaying function for Scheme b, and a non-

monotonic function for Scheme c. In the model calculation, the mean time elapsed during product creation is set equal to $0.5\,\gamma^{-1}$ for all cases. The solid lines represent the predictions of Eq. (3) for Scheme a, Scheme b with $l=2$, and Scheme c with $l=20$. Simulation results are represented by the squares, triangles, and circles. For more detailed information about the models and simulation method, see Supplemental Material.

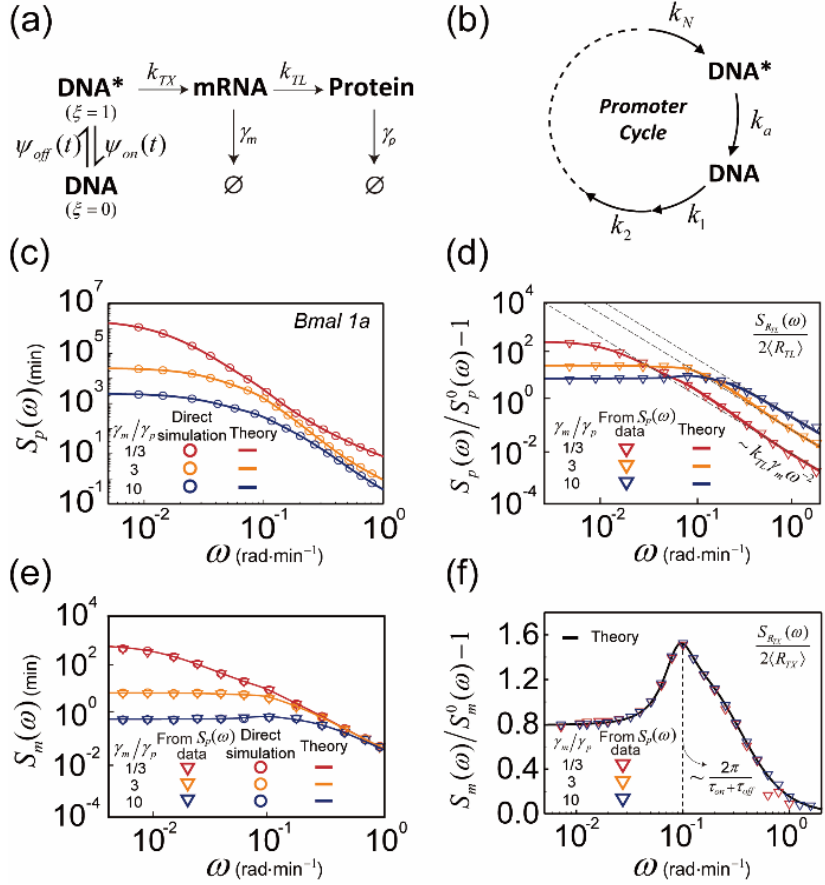

**Fig. 2. Spectral analysis of the protein number fluctuation.** (a) Reaction network model of *luciferase* expression controlled by the *Bmal 1a* promoter in mouse fibroblast cells [14]. $\psi_{on}(t)$ and $\psi_{off}(t)$ represent the lifetime distribution of the active and inactive gene states, respectively. (b) Model of the gene deactivation and activation cycle of the *Bmal 1a* promoter. Gene deactivation is a simple Poisson process, but gene activation is a non-Poisson process composed of *N* consecutive Poisson processes. (c) Power spectrum of protein number at three different ratios of the mRNA lifetime to the protein lifetime. (circle) simulation results; (lines) theoretical results. (d) Modified power spectrum of protein number or mean-scaled power spectrum of translation rate, $S_p(\omega)/S_p^0(\omega)-1 \left[ = S_{R_{TL}}(\omega)/2\langle R_{TL}\rangle \right]$. $S_p^0(\omega)$ is defined as $S_p^0(\omega) = 2\langle R\rangle/(\omega^2 + \gamma_p^2)$. In the high frequency regime, the asymptotic behavior of the modified power spectrum is given by $k_{TL}\gamma_m\omega^{-2}$, i.e., $\lim_{\omega\to\infty}\omega^2\left[S_p(\omega)/S_p^0(\omega)-1\right]=k_{TL}\gamma_m$. (e) Power spectrum of the mRNA number. (triangles) the data extracted from the power spectrum data, $S_p(\omega)$, of the protein number with use of Eq. (6). (circles) simulation results, (lines) the theoretical results of Eqs. (3) and (8). (f) Modified mRNA number power-spectrum or mean-scaled power spectrum of transcription rate. (triangles) the data extracted from $S_p(\omega)$ with use of Eqs. 6 and 7. (line) the theoretical result of Eq. (8).

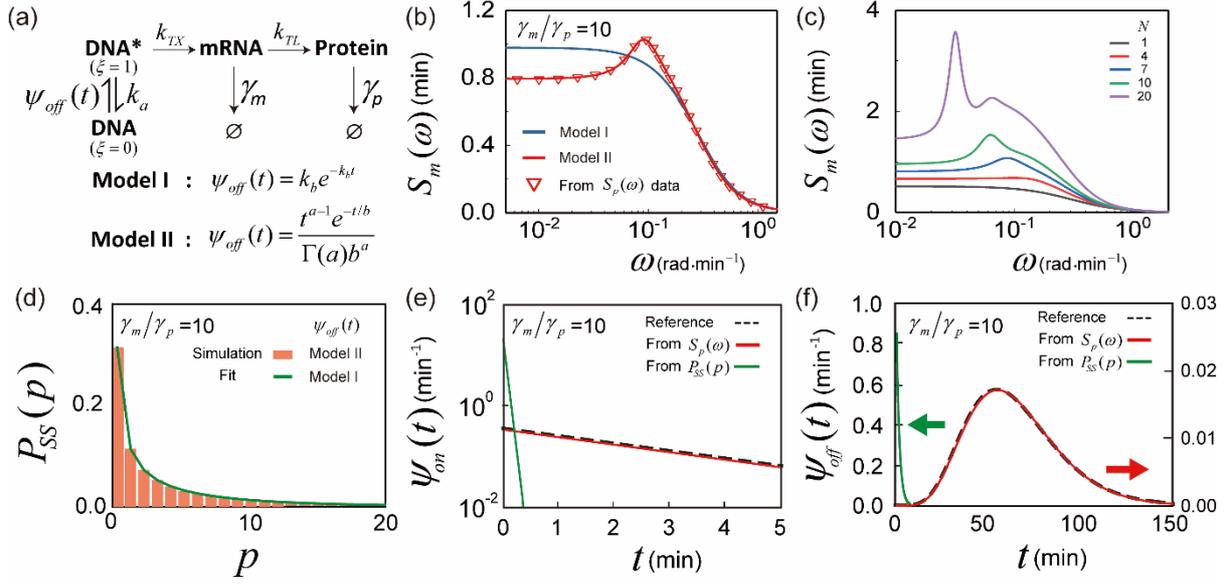

**Fig. 3. Quantitative analysis of the protein number power spectrum.** (a) Gene expression network models with two different gene activation processes: Poisson process (Model I) and non-Poisson process with a gamma distributed waiting time (Model II). When the gene activation process composed of $N$ consecutive Poisson processes with the rate constants being given by the same constant, $k$, the lifetime distribution, $\psi_{off}(t)$, of the inactive gene state is given by a gamma distribution, $t^{a-1}e^{-t/b}/b^a\Gamma(a)$, with $a$ and $b$ being $N$ and $k^{-1}$, respectively. (b) (triangles) mRNA number power spectrum corresponding to the data shown by the blue triangles in Fig. 3f. (blue line) the best fitted result of Model I, (red line) the best fitted result of Model II. (c) Dependence of $S_m(\omega)$ on the number, $N$, of the consecutive reaction steps comprising the gene activation process. As $N$ increases, the peak position shifts to the lower frequency, and the peak shape gets sharper. (d) Steady-state distribution of the protein number. (histogram) Simulation results for the gene expression network model in Fig. 3a and b. (green line) the best fitted result of the incorrect model, Model I. (circles) (e, f) Lifetime distributions, $\psi_{on}(t)$ and $\psi_{off}(t)$, of the active and inactive gene states, extracted

from the power spectrum analysis by Model I (blue line) and Model II (red line), and from the steady-state protein number distribution by Model I (green line). $\psi_{on}(t)$ and $\psi_{off}(t)$, which were extracted from the power spectrum analysis with use of Model II, are in excellent agreement with the reference model in Fig. 2a and b. The steady state distribution of the protein number in Fig. 3d can be explained by Model I (green line). However, $\psi_{on}(t)$ and $\psi_{off}(t)$, extracted from the steady-state protein number distribution by Model I, show a great deal of deviation from the correct results.

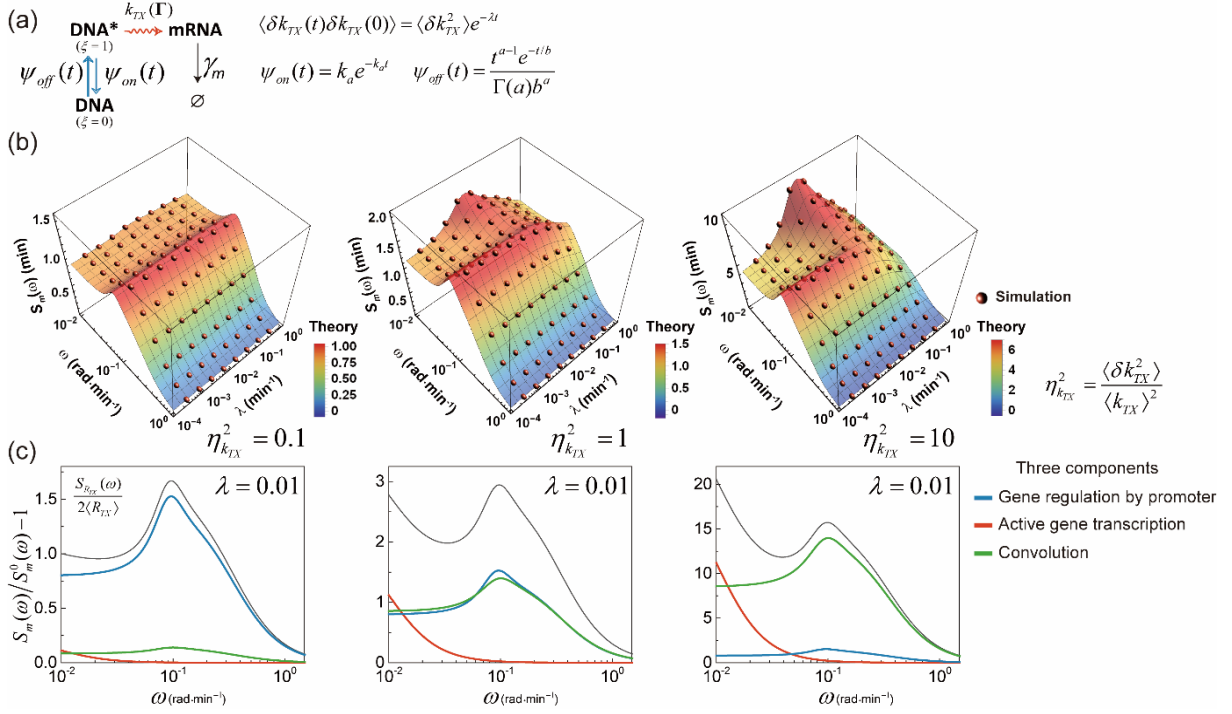

**Fig. 4. mRNA number power spectrum of gene expression network model with the super-Poisson active gene transcription dynamics.** (a) Gene expression network model with the *vibrant* active gene transcription process. The active gene transcription rate, $k_{TX}$, is a dynamic, stochastic variable. The mean value, $\langle k_{TX} \rangle$, of the active gene transcription rate is the same as the value of $k_{TX}$ in Figs. 3 and 4. The normalized time correlation function (TCF) of $k_{TX}$ is given by an exponentially decaying function of time, i.e., $\phi_{k_{TX}}(t) = \exp(-\lambda t)$. An example of a reaction scheme with the exponential TCF of the rate fluctuation is a two-channel reaction model (see Fig. 1b and f). Gene regulation by the promoter is modeled the same as Model II in Fig. 3b. (b) Dependence of the mRNA number power spectrum on the relaxation rate, $\lambda$, of the fluctuation in $k_{TX}$ for three different magnitudes of $k_{TX}$ fluctuation. $\eta^2_{k_{TX}}$ denotes the relative variance of $k_{TX}$. (surface) The theoretical result for the mRNA number power

spectrum calculated by Eqs. (3) and (9). (spheres) The stochastic simulation result (see Supplemental Material, Method 8). The mRNA number power spectrum has a peak around the promoter fluctuation frequency $\omega_{peak} = 2\pi/(\tau_{on} + \tau_{off}) \cong 0.1$ rad·s$^{-1}$, originating from the sub-Poisson gene regulation dynamics by the promoter. The peak diminishes when $\eta_{k_{TX}}^2$ is large and $k_{TX}$ fluctuates fast. (c) Three components in the power spectrum of the transcription rate. (black line) the value of $S_{R_{TX}}(\omega)/(2\langle R_{TX} \rangle) \left[ = \langle R_{TX} \rangle \tilde{S}_{R_{TX}}(\omega)/2 \right]$ or $S_m(\omega)/S_m^0(\omega) - 1$ with $S_m^0(\omega) = 2(\langle m \rangle \gamma_m)/(\omega^2 + \gamma_m^2)$. The value of $\lambda$ is set to be 0.01 min$^{-1}$. (colored lines) three components of the power-spectrum: $S_\xi(\omega)$ originating from the gene regulating dynamics of the promoter (blue), $S_{k_{TX}}(\omega)$ originating from the active gene transcription dynamics (red), and their convolution $S_\xi(\omega) * S_{k_{TX}}(\omega)$ (green). See Eq. (9). The relative contribution of $S_{k_{TX}}(\omega)$ and $S_\xi(\omega) * S_{k_{TX}}(\omega)$ increases with $\eta_{k_{TX}}^2$, while $S_\xi(\omega)$ does not. When the value of $\eta_{k_{TX}}^2$ is 10, the relative contribution of $S_\xi(\omega)$ is marginal, but the non-monotonic peak in the power spectrum persists due to the contribution of the convolution term, $S_\xi(\omega) * S_{k_{TX}}(\omega)$, given that $\lambda$ is smaller than $\omega_{peak}$ (see Fig. S2 for the case where $\lambda$ is as large as $\omega_{peak}$).

# Supporting Information for

# Frequency spectrum of intracellular chemical noise: a sensitive probe of the reaction dynamics in living cells

Sanggeun Song[1-3], Gil-Suk Yang[1-3], Seong Jun Park[1-3], Ji-Hyun Kim[1*], and Jaeyoung Sung[1-3*]

[1]Creative Research Initiative Center for Chemical Dynamics in Living Cells, Chung-Ang University, Seoul 06974, Korea.

[2]Department of Chemistry, Chung-Ang University, Seoul 06974, Korea.

[3]National Institute of Innovative Functional Imaging, Chung-Ang University, Seoul 06974, Korea.

**Author Information** The authors declare no competing financial interests. Correspondence and requests for materials should be addressed to J.S. (jaeyoung@cau.ac.kr) and J.-H.K. (jihyunkim@cau.ac.kr).

# Table of Contents

## I. SUPPLEMENTARY METHODS

## II. SUPPLEMENTARY NOTES

## III. SUPPLEMENTARY FIGURES

## IV. SUPPLEMENTARY REFERENCES

# I. SUPPLEMENTARY METHODS

**Supplementary Method 1| Derivation of Eq. 3.**

In this section, we provide a detailed derivation of Eq. 3, one of the key results in this work. First, we consider the Fourier transform of the product number time autocorrelation function, given on the right-hand side (R.H.S) of Eq. 2. Using the Euler's formula, $e^{i\theta} = \cos\theta + i\sin\theta$, we can write the Fourier transform as

$$\int_{-\infty}^{\infty} dt\, e^{-i\omega t} \langle \delta z(t)\delta z(0) \rangle_{ss} = \int_{-\infty}^{\infty} dt\, \cos(\omega t)\langle \delta z(t)\delta z(0) \rangle_{ss} - i\int_{-\infty}^{\infty} dt\, \sin(\omega t)\langle \delta z(t)\delta z(0) \rangle_{ss},$$

[M1-1]

which is known as the cosine and sine transform. In the steady-state, the time correlation function of product number, $\langle \delta z(t)\delta z(0) \rangle_{ss}$, is an even function, that is, $\langle \delta z(-t)\delta z(0) \rangle_{ss} = \langle \delta z(t)\delta z(0) \rangle_{ss}$, so that the first term on the R.H.S as

$$\int_{-\infty}^{\infty} dt\, \cos(\omega t)\langle \delta z(t)\delta z(0) \rangle_{ss} = 2\operatorname{Re}\left[\int_{0}^{\infty} dt\, e^{-i\omega t}\langle \delta z(t)\delta z(0) \rangle_{ss}\right],$$

[M1-2]

and the second term on the R.H.S vanishes. Recalling the definition of the Laplace transform, the R.H.S of the above equation is same as the Laplace transform of $\langle \delta z(t)\delta z(0) \rangle_{ss}$, where the Laplace variable is equal to $i\omega$. When we denote the Laplace transform of the product number time autocorrelation function as $\hat{\varphi}_z(s) \equiv \int_{0}^{\infty} dt\, e^{-st}\langle \delta z(t)\delta z(0) \rangle_{ss}$, the relationship between the power spectrum of the product number satisfies

$$S_z(\omega) = 2\operatorname{Re}[\hat{\varphi}_z(i\omega)] = \hat{\varphi}_z(i\omega) + \hat{\varphi}_z(-i\omega).$$

[M1-3]

To derive the expression of $\hat{\varphi}_z(s)$ for an intracellular birth-death process, where the product creation rate is coupled to cell state variables, we consider the following generalized master equation (GME)(1),(2)

$$\frac{\partial}{\partial t} p_z(\Gamma,t) = R(\Gamma)[p_{z-1}(\Gamma,t) - p_z(\Gamma,t)] + \gamma[(z+1)p_{z+1}(\Gamma,t) - zp_z(\Gamma,t)] + L(\Gamma)p_z(\Gamma,t),$$

[M1-4]

where $p_z(\Gamma,t)$ is the probability that $z$ number of products exist at cell state $\Gamma$ at time $t$. $R(\Gamma)$ denotes the product creation rate dependent on the cell state variable, and $L(\Gamma)$ designates the time evolution operator of cell state variable, $\Gamma$. The steady-state time correlation function (TCF) of the product number is given by

$$\langle z(t)z(0)\rangle_{ss} = \int d\Gamma_0 \sum_{z_0=0}^{\infty} \langle z(t|z_0,\Gamma_0)\rangle z_0 p_{ss}(z_0,\Gamma_0),$$

[M1-5]

where $\langle z(t|z_0,\Gamma_0)\rangle$ and $p_{ss}(z_0,\Gamma_0)$ respectively denote the average number of the product molecules at time $t$, given that the product number is given by $z_0$ and the cell state is at $\Gamma_0$ at time 0, and the joint distribution function of the product number and the cell state variable in the steady-state. The mathematical definition of $\langle z(t|z_0,\Gamma_0)\rangle$ is given by $\langle z(t|z_0,\Gamma_0)\rangle \equiv \int d\Gamma \sum_{z=0}^{\infty} zp_z(\Gamma,t|z_0,\Gamma_0)$, where $p_z(\Gamma,t|z_0,\Gamma_0)$ denotes the solution of Eq. M1-4 under the following initial condition, $p_z(\Gamma,t=0) = \delta_{zz_0}\delta(\Gamma-\Gamma_0)$. To obtain $\langle z(t|z_0,\Gamma_0)\rangle$, let us first obtain the time-evolution equation for $\langle z(\Gamma,t|z_0,\Gamma_0)\rangle$, or $\sum_{z=0}^{\infty} zp_z(\Gamma,t|z_0,\Gamma_0)$, the integration of which over $\Gamma$ yields $\langle z(t|z_0,\Gamma_0)\rangle$. From Eq. M1-4, we obtain

$$\frac{\partial}{\partial t}\langle z(\Gamma,t\,|\,z_0,\Gamma_0)\rangle = R(\Gamma)p(\Gamma,t\,|\,z_0,\Gamma_0) - \gamma\langle z(\Gamma,t\,|\,z_0,\Gamma_0)\rangle + L(\Gamma)\langle z(\Gamma,t\,|\,z_0,\Gamma_0)\rangle \quad ,$$

[M1-6]

where $p(\Gamma,t\,|\,z_0,\Gamma_0)$ is defined as $p(\Gamma,t\,|\,z_0,\Gamma_0) \equiv \sum_{z=0}^{\infty} p_z(\Gamma,t\,|\,z_0,\Gamma_0)$. By taking the Laplace transform of Eq. M1-5, we obtain the following formal expression for the Laplace transform of $\langle z(\Gamma,t\,|\,z_0,\Gamma_0)\rangle$:

$$\langle \hat{z}(\Gamma,s\,|\,z_0,\Gamma_0)\rangle = \left[s+\gamma-L(\Gamma)\right]^{-1}\left[z_0\delta(\Gamma-\Gamma_0) + R(\Gamma)\hat{p}(\Gamma,s\,|\,z_0,\Gamma_0)\right]. \qquad [\text{M1-7}]$$

To obtain this equation, we have used the following initial condition, $\langle z(\Gamma,0\,|\,z_0,\Gamma_0)\rangle = \sum_{z=z_0}^{\infty} z\delta_{zz_0}\delta(\Gamma-\Gamma_0) = z_0\delta(\Gamma-\Gamma_0)$. In terms of the Green's function defined by $\hat{G}(\Gamma,s\,|\,\Gamma_0) = \left[s-L(\Gamma)\right]^{-1}\delta(\Gamma-\Gamma_0)$, we can rewrite Eq. M1-6 as follows:

$$\langle \hat{z}(\Gamma,s\,|\,z_0,\Gamma_0)\rangle = z_0\hat{G}(\Gamma,s+\gamma\,|\,\Gamma_0) + \left[s+\gamma-L(\Gamma)\right]^{-1}R(\Gamma)\hat{p}(\Gamma,s\,|\,z_0,\Gamma_0). \qquad [\text{M1-8}]$$

Our next task is to find the expression for $p(\Gamma,t\,|\,z_0,\Gamma_0)\left[\equiv \sum_{z=0}^{\infty} p_z(\Gamma,t\,|\,z_0,\Gamma_0)\right]$, whose Laplace transform appears on the R.H.S. of Eq. M1-7. By applying $\sum_{z=0}^{\infty}(\cdots)$ on both sides of Eq. M1-4, we obtain the evolution equation of $p(\Gamma,t\,|\,z_0,\Gamma_0)$ as follows:

$$\frac{\partial}{\partial t}p(\Gamma,t\,|\,z_0,\Gamma_0) = L(\Gamma)p(\Gamma,t\,|\,z_0,\Gamma_0). \qquad [\text{M1-9}]$$

Noting that the initial condition of this equation is given by $p(\Gamma,0\,|\,\Gamma_0) = \delta(\Gamma-\Gamma_0)$, one can show that the solution of Eq. M1-8 is actually the same as the Green's function defined above

in Eq. M1-7, that is, $\hat{p}(\Gamma,s|z_0,\Gamma_0) = [s-L(\Gamma)]^{-1}\delta(\Gamma-\Gamma_0) = \hat{G}(s,\Gamma|\Gamma_0)$. Substituting the latter equation into Eq. M1-7, and using the following identity,

$$[s+\gamma-L(\Gamma)]^{-1}f(\Gamma) = [s+\gamma-L(\Gamma)]^{-1}\int d\Gamma\,\delta(\Gamma-\Gamma_1)f(\Gamma_1)$$
$$= \int d\Gamma_1 \hat{G}(\Gamma,s+\gamma|\Gamma_1)f(\Gamma_1)$$
, [M1-10]

we obtain

$$\langle\hat{z}(\Gamma,s|z_0,\Gamma_0)\rangle = z_0\hat{G}(\Gamma,s+\gamma|\Gamma_0) + \int d\Gamma_1 \hat{G}(\Gamma,s+\gamma|\Gamma_1)R(\Gamma_1)\hat{G}(\Gamma_1,s|\Gamma_0).$$ [M1-11]

Taking the integral over $\Gamma$ on both sides of this equation, we obtain

$$\langle\hat{z}(s|z_0,\Gamma_0)\rangle = \frac{z_0}{s+\gamma} + \frac{1}{s+\gamma}\int d\Gamma\,R(\Gamma)\hat{G}(\Gamma,s|\Gamma_0).$$ [M1-12]

In the derivation of Eq. M1-12, we have used the following identity, $\int d\Gamma\,\hat{G}(\Gamma,s|\Gamma_0) = 1/s$, which is nothing but the Laplace transform of the normalization condition, i.e., $\int d\Gamma\,G(\Gamma,t|\Gamma_0) = 1$. Substituting Eq. M1-12 into the Laplace transform of Eq. M1-5, we obtain

$$\mathcal{L}[\langle z(t)z(0)\rangle_{ss}] = \sum_{z_0=0}^{\infty}\int d\Gamma_0 \langle\hat{z}(s|z_0,\Gamma_0)\rangle z_0 p_{ss}(z_0,\Gamma_0)$$
$$= \int d\Gamma_0 \sum_{z_0=0}^{\infty} \frac{z_0^2}{s+\gamma} p_{ss}(z_0,\Gamma_0)$$
$$+ \int d\Gamma_0 \sum_{z_0=0}^{\infty} \frac{z_0}{s+\gamma}\int d\Gamma\,R(\Gamma)\hat{G}(\Gamma,s|\Gamma_0)p_{ss}(z_0,\Gamma_0)$$
. [M1-13]

Rearranging the above equation, we find that

$$\mathcal{L}[\langle z(t)z(0)\rangle_{ss}] = \frac{\langle z^2\rangle_{ss}}{s+\gamma} + \frac{1}{s+\gamma}\int d\Gamma\int d\Gamma_0 R(\Gamma)\hat{G}(\Gamma,s|\Gamma_0)\langle z(\Gamma_0)\rangle_{ss},$$ [M1-14]

where $\langle z^2 \rangle_{ss}$ and $\langle z(\Gamma_0) \rangle_{ss}$ are given by $\int d\Gamma_0 \sum_{z_0=0}^{\infty} z_0^2 p_{ss}(z_0, \Gamma_0) = \sum_{z_0=0}^{\infty} z_0^2 p_{ss}(z_0)$ and $\sum_{z_0=0}^{\infty} z_0 p_{ss}(z_0, \Gamma_0)$, respectively. To obtain the expression for $\langle z(\Gamma_0) \rangle_{ss}$, we should obtain the expression for $p_{ss}(z, \Gamma)$. Due to the fact that, in the steady-state, the probability density satisfying Eq. M1-4 does not change over time, we obtain the following equation for $p_{ss}(z, \Gamma)$:

$$R(\Gamma)[p_{ss}(z-1, \Gamma) - p_{ss}(z, \Gamma)] - \gamma(z+1)p_{ss}(z+1, \Gamma) - zp_{ss}(z, \Gamma) + L(\Gamma)p_{ss}(z, \Gamma) = 0$$

[M1-15]

By applying $\sum_{z=0}^{\infty} z$ on both sides of this equation and then rearranging the resulting equation, we can obtain

$$\begin{aligned}\langle z(\Gamma) \rangle_{ss} &= [\gamma - L(\Gamma)]^{-1} R(\Gamma) p_{ss}(\Gamma) \\ &= \int d\Gamma_1 \hat{G}(\Gamma, \gamma | \Gamma_1) R(\Gamma_1) p_{ss}(\Gamma_1)\end{aligned}$$

[M1-16]

Substituting Eq. M1-16 into Eq. M1-14, we obtain

$$\begin{aligned}\mathcal{L}[\langle z(t)z(0) \rangle_{ss}] &= \frac{\langle z^2 \rangle_{ss}}{s+\gamma} + \frac{1}{s+\gamma} \int d\Gamma \int d\Gamma_0 R(\Gamma) \hat{G}(\Gamma, s | \Gamma_0)[\gamma - L(\Gamma_0)]^{-1} R(\Gamma_0) p_{ss}(\Gamma_0) \\ &= \frac{\langle z^2 \rangle_{ss}}{s+\gamma} + \frac{1}{s+\gamma} \int d\Gamma \int d\Gamma_0 R(\Gamma)[s - L(\Gamma)]^{-1} \delta(\Gamma - \Gamma_0)[\gamma - L(\Gamma_0)]^{-1} R(\Gamma_0) p_{ss}(\Gamma_0) \\ &= \frac{\langle z^2 \rangle_{ss}}{s+\gamma} + \frac{1}{s+\gamma} \int d\Gamma R(\Gamma)[s - L(\Gamma)]^{-1}[\gamma - L(\Gamma)]^{-1} R(\Gamma) p_{ss}(\Gamma)\end{aligned}$$

.                                            [M1-17]

Taking advantage of the following operator identity, $A^{-1}B^{-1} = \frac{1}{A-B}(B^{-1} - A^{-1})$, where $A - B$ is a scalar function, we have $[s - L(\Gamma)]^{-1}[\gamma - L(\Gamma)]^{-1} = (s - \gamma)^{-1}\left\{[\gamma - L(\Gamma)]^{-1} - [s - L(\Gamma)]^{-1}\right\}$.

Substituting this identity into Eq. M1-17, we obtain

$$\mathcal{L}[\langle z(t)z(0)\rangle_{ss}] = \frac{\langle z^2\rangle_{ss}}{s+\gamma} - \frac{1}{s^2-\gamma^2}\int d\Gamma R(\Gamma)[s-L(\Gamma)]^{-1}R(\Gamma)p_{ss}(\Gamma)$$
$$+\frac{1}{s^2-\gamma^2}\int d\Gamma R(\Gamma)[\gamma-L(\Gamma)]^{-1}R(\Gamma)p_{ss}(\Gamma)$$
[M1-18]

In terms of Green's function, $\int d\Gamma R(\Gamma)[s-L(\Gamma)]^{-1}R(\Gamma)p_{ss}(\Gamma)$, appearing in Eq. M1-18, can be written as $\int d\Gamma \int d\Gamma_0 R(\Gamma)\hat{G}(\Gamma,s|\Gamma_0)R(\Gamma_0)p_{ss}(\Gamma_0)$, and this is nothing but the Laplace transform of the steady-state time correlation function (TCF) of the product creation rate defined by

$$\langle R(t)R(0)\rangle_{ss} \equiv \int d\Gamma \int d\Gamma_0 R(\Gamma)G(t,\Gamma|\Gamma_0)R(\Gamma_0)p_{ss}(\Gamma_0). \quad [\text{M1-19}]$$

In terms of the TCF, Eq. M1-18 can be written as

$$\mathcal{L}[\langle z(t)z(0)\rangle_{ss}] = \frac{\langle z^2\rangle_{ss}}{s+\gamma} + \frac{1}{s^2-\gamma^2}\left[\int_0^\infty dt e^{-\gamma t}\langle R(t)R(0)\rangle - \int_0^\infty dt e^{-st}\langle R(t)R(0)\rangle\right] \quad [\text{M1-20}]$$

Subtracting $\langle z\rangle_{ss}^2/s$ from the both sides of Eq. M1-20, and using the following identities, $\langle z\rangle_{ss} = \langle R\rangle/\gamma$ and $\langle R(t)R(0)\rangle = \langle \delta R(t)\delta R(0)\rangle + \langle R\rangle^2$, we obtain

$$\mathcal{L}[\langle \delta z(t)\delta z(0)\rangle_{ss}] = \frac{\langle \delta z^2\rangle_{ss}}{s+\gamma} + \frac{1}{s^2-\gamma^2}\left[\int_0^\infty dt e^{-\gamma t}\langle \delta R(t)\delta R(0)\rangle - \int_0^\infty dt e^{-st}\langle \delta R(t)\delta R(0)\rangle\right],$$
$$= \frac{\langle \delta z^2\rangle_{ss}}{s+\gamma} + \frac{\langle \delta R\rangle^2}{s^2-\gamma^2}\left[\hat{\phi}_R(\gamma)-\hat{\phi}_R(s)\right]$$

[M1-21]

with $\hat{\phi}_q(s) = \int_0^\infty dt e^{-st}\langle \delta q(t)\delta q(0)\rangle / \langle \delta q^2\rangle$. By dividing both sides of Eq. M1-21 by $\langle \delta z^2\rangle_{ss}$, we obtain

$$\hat{\phi}_z(s) = \frac{1}{s+\gamma} + \frac{\eta_R^2}{\eta_z^2}\frac{\gamma^2}{s^2-\gamma^2}\left[\hat{\phi}_R(\gamma) - \hat{\phi}_R(s)\right], \qquad [\text{M1-22}]$$

where $\eta_z^2$ and $\eta_R^2$ are, respectively, the relative variance of the product number, $\eta_z^2 = \langle \delta z^2 \rangle / \langle z \rangle^2$ and its rate, $\eta_R^2 = \langle \delta R^2 \rangle / \langle R \rangle^2$. This result has been previously reported in Eq. (D5) in Ref. (2).

Noting that the Laplace transform of $\langle z(t)z(0)\rangle_{ss}$, or $\hat{\varphi}(s)$, is given by $\langle \delta z^2 \rangle \hat{\phi}_z(s)$, we can obtain the expression for the power spectrum of the product number from Eq. M1-3 and Eq. M1-22 as follows:

$$\begin{aligned} S_z(\omega) &= \langle \delta z^2 \rangle [\hat{\phi}_z(i\omega) + \hat{\phi}_z(-i\omega)] \\ &= \frac{2\langle \delta z^2 \rangle \gamma}{\omega^2 + \gamma^2}\left[1 - \frac{\eta_R^2}{\eta_z^2}\gamma\hat{\phi}_R(\gamma)\right] + \frac{1}{\omega^2 + \gamma^2}\frac{\langle z \rangle^2}{\langle R \rangle^2}S_R(\omega) \end{aligned} \qquad [\text{M1-23}]$$

Here, $S_R(\omega)$ denotes the power spectrum of product creation rate, i.e., $S_R(\omega) = \langle \delta R^2 \rangle \left[\hat{\phi}_R(i\omega) - \hat{\phi}_R(-i\omega)\right]$. Noting that $\langle \delta z^2 \rangle \gamma^2 \eta_R^2 / \eta_z^2 = \langle z \rangle^2 \gamma^2 \eta_R^2 = \langle R \rangle^2 \eta_R^2 = \langle \delta R^2 \rangle$ and $\langle z \rangle^2 = (\langle R \rangle / \gamma)^2$, we can simplify Eq. M1-23 to

$$S_z(\omega) = \frac{2}{\omega^2 + \gamma^2}\left[\gamma \langle \delta z^2 \rangle - \langle \delta R^2 \rangle \hat{\phi}_R(\gamma)\right] + \frac{S_R(\omega)}{\omega^2 + \gamma^2}. \qquad [\text{M1-24}]$$

It is known that, for the general birth-death process, the Fano factor, $F_z (\equiv \langle \delta z^2 \rangle / \langle z \rangle = \eta_z^2 \langle z \rangle)$, of the product noise obeys the following equation:

$$F_z - 1 = \hat{\phi}_R(\gamma) F_R \qquad [\text{M1-25}]$$

By multiplying $\langle z \rangle (= \langle R \rangle / \gamma)$ on both sides of Eq. M1-25, we obtain $\langle \delta z^2 \rangle = \langle z \rangle + \hat{\phi}_R(\gamma) \langle \delta R^2 \rangle / \gamma$. Substituting this equation into Eq. M1-24, we finally obtain

$$S_z(\omega) = \frac{2\langle R \rangle}{\omega^2 + \gamma^2} + \frac{S_R(\omega)}{\omega^2 + \gamma^2}, \qquad \text{[M1-26]}$$

which is Eq. 3 in the main text.

**Supplementary Method 2 | Derivation of Eq. 5.**

In this Method, we will begin by deriving the exact relationship between the time correlation function (TCF) of the product creation rate and the time series of reaction events, or the series of times at which reaction event occurs. Let us consider a time series, $\{t_1, t_2, \cdots\}$, of reaction events, where $t_i$ denotes the time at which the $i$-th reaction event is completed and the $i$-th product molecule is created. The number, $z(t)$, of reaction events occurring in time interval $(0, t)$ is given by $\sum_{i=1}^{\infty} \Theta(t - t_i)$, where $\Theta(x)$ denotes the Heaviside step function. As the chemical reaction process is a stochastic process, the reaction times $\{t_i\}$ and $z(t)$ are random variables. The rate $R$ of the reaction is defined as $R = dz(t)/dt$. Noting that $d\Theta(t - t_i)/dt = \delta(t - t_i)$, where $\delta(x)$ is the Dirac delta function, we obtain

$$R(t) = \sum_{i=1}^{\infty} \delta(t - t_i). \qquad [\text{M2-1}]$$

The time correlation function of the reaction rate, $\langle R(\tau_2) R(\tau_1) \rangle$, is defined by

$$\langle R(\tau_2) R(\tau_1) \rangle = \sum_{i=1}^{\infty} \sum_{\substack{j=1 \\ j \neq i}}^{\infty} \langle \delta(\tau_2 - t_i) \delta(\tau_1 - t_j) \rangle. \qquad [\text{M2-2}]$$

Because $t_j$ is the time at which the $j$-th reaction event is completed, $t_j$ increases with $j$. With this notation, one can obtain the following equation from Eq. M2-2,

$$\langle R(t + t_0) R(t_0) \rangle = \sum_{i=1}^{\infty} \sum_{j=i+1}^{\infty} \langle \delta(t + t_0 - t_j) \delta(t_0 - t_i) \rangle \quad (t > 0), \qquad [\text{M2-3}]$$

by noting that $\langle \delta(t+t_0 - t_j)\delta(t_0 - t_i)\rangle = 0$ for all $j$ less than $i$ when $t > 0$. This equation simply means that if the $i$-th reaction events occurs at $t_0$ and the $j$-th reaction event occurs at a later time $t + t_0$ $(t > 0)$, $j$ should be greater than $i$ for any time series of reaction events. Using the following property of the Dirac delta function, $f(t_0)\delta(t_0 - t_i) = f(t_i)\delta(t_0 - t_i)$, one can rewrite Eq. M2-3 as

$$\langle R(t+t_0)R(t_0)\rangle = \sum_{i=1}^{\infty}\sum_{j=i+1}^{\infty} \langle \delta(t-(t_j - t_i))\delta(t_0 - t_i)\rangle$$
$$= \sum_{i=1}^{\infty}\sum_{l=1}^{\infty} \langle \delta(t-(t_{l+i} - t_i))\delta(t_0 - t_i)\rangle$$

[M2-4]

Let us confine ourselves to the case where our reaction process is a stationary process, for which $t_{i+l} - t_i$, or the magnitude of the time interval between the $i$-th reaction event and the $(i+l)$-th reaction event, is independent of the time $t_i$ at which the $i$-th reaction event occurs. Therefore, we can rewrite Eq. M2-4 as

$$\langle R(t+t_0)R(t_0)\rangle = \sum_{i=1}^{\infty}\sum_{l=1}^{\infty} \langle \delta(t-(t_{l+i} - t_i))\rangle\langle \delta(t_0 - t_i)\rangle$$
$$= \sum_{i=1}^{\infty}\langle \delta(t_0 - t_i)\rangle \sum_{l=1}^{\infty}\langle \delta(t-(t_{l+i} - t_i))\rangle \quad ,$$
$$= \langle R \rangle \sum_{l=1}^{\infty}\langle \delta(t-(t_{l+i} - t_i))\rangle$$

[M2-5]

for a stationary reaction process, for which the average, $\langle R(t)\rangle$, of the reaction rate over a large number of the reaction time series is constant in time, i.e., $\langle R(t)\rangle = \langle R \rangle$ at any $t$. In Eq. M2-5, $\langle \delta(t-(t_{l+i} - t_i))\rangle$ is nothing but the probability density, $\psi_l(t)$, of the time, $t = t_{i+l} - t_i$, elapsed from the $i$-th reaction event to the $(i+l)$-th reaction event, or the probability density of the time required for $l$ reaction events to take place since the $i$-th reaction event takes place at $t_i$.

From its definition, it is obvious that $\psi_l(t)$ satisfies the normalization condition, $\int_0^\infty dt \psi_l(t) = 1$, for any $l$. In terms of $\psi_l(t)$, Eq. M2-5 can be rewritten as

$$\langle R(t+t_0)R(t_0)\rangle = \langle R\rangle \sum_{l=1}^\infty \psi_l(t) \qquad \text{[M2-6]}$$

Noting that the correlation between $R(t+t_0)$ and $R(t_0)$ vanishes in the long time limit, i.e., $\lim_{t\to\infty}\langle R(t+t_0)R(t_0)\rangle = \langle R\rangle^2$, one can obtain the following relation between the mean reaction rate and $\psi_l(t)$ in the long time limit:

$$\lim_{t\to\infty} \sum_{l=1}^\infty \psi_l(t) = \langle R\rangle \qquad \text{[M2-7]}$$

From Eq. M2-6, we obtain

$$\begin{aligned}\langle \delta R(t+t_0)\delta R(t_0)\rangle &= \langle R(t+t_0)R(t_0)\rangle - \langle R\rangle^2 \\ &= \langle R\rangle \left[\sum_{l=1}^\infty \psi_l(t) - \langle R\rangle\right]\end{aligned} \qquad \text{[M2-8]}$$

By dividing Eq. M2-8 by $\langle R\rangle$ and taking the Laplace transform on both sides of the resulting equation, we obtain

$$F_R \hat{\phi}_R(s) = \sum_{l=0}^\infty \hat{\psi}_l(s) - \frac{\langle R\rangle}{s}, \qquad \text{[M2-9]}$$

where $F_R$ is the Fano factor of the creation rate, $F_R = \langle \delta R^2\rangle/\langle R\rangle$. Substituting Eq. M2-9 into M1-3, or $S_R(\omega) = 2\langle R\rangle F_R \lim_{\varepsilon\to 0^+} \text{Re}\left[\hat{\phi}_R(\varepsilon + i\omega)\right]$, we can easily obtain the expression for the power spectrum of the product creation rate.

For the special case when the product creation process in a renewal process, we have $\hat{\psi}_l(s) = \hat{\psi}_1(s)^l$, so that Eq. M2-9 becomes

$$F_R \hat{\phi}_R(s) = \frac{\hat{\psi}_1(s)}{1-\hat{\psi}_1(s)} - \frac{\langle R \rangle}{s} \ . \qquad [\text{M2-10}]$$

Substituting Eq. M2-9 into Eq. M1-3, or $S_R(\omega) = 2\langle R \rangle F_R \lim_{\varepsilon \to 0^+} \text{Re}\left[\hat{\phi}_R(\varepsilon + i\omega)\right]$, we obtain the following expression for the power spectrum, $S_R(\omega)$, of the product creation rate for a renewal product creation process:

$$S_R(\omega) = 2\langle R \rangle \text{Re}\left[\frac{\hat{\psi}_1(i\omega)}{1-\hat{\psi}_1(i\omega)}\right] - 2\langle R \rangle^2 \lim_{\varepsilon \to 0^+} \frac{\varepsilon}{\varepsilon^2 + \omega^2} \qquad [\text{M2-11}]$$

Noticing that $\psi_1(t)$ is zero when $t < 0$, we define the Fourier transform of $\psi_1(t)$ as $\tilde{\psi}_1(\omega) = \int_0^\infty dt\, e^{-i\omega t} \psi_1(t)$. Then, the above equation can also be written as

$$S_R(\omega) = 2\langle R \rangle \text{Re}\left[\frac{\tilde{\psi}(\omega)}{1-\tilde{\psi}(\omega)}\right] - 2\pi \langle R \rangle^2 \delta(\omega) \qquad [\text{M2-12}]$$

For nonzero frequency $(\omega > 0)$, Eq. M2-12 is the same as Eq. 5 in the main text.

**Supplementary Method 3 | Power spectrum of the product creation rate for multi-channel and multi-step processes.**

In this section, we briefly derive the power spectrum of the product creation rate for the multi-channel reaction process and the multi-step process, shown in Fig. 1B and C. from the main text. For the *l*-channel process shown in Fig. 1B, the time correlation function of the product creation rate can be written as

$$\langle R(t)R(0) \rangle_{ss} = \sum_{i=1}^{l} \sum_{j=1}^{l} R_j G(\Gamma_j, t | \Gamma_i) R_i P_{ss}(\Gamma_i) \qquad \text{[M3-1]}$$

which is equivalent to Eq. M1-19 for the case where the state variable is discrete. In Eq. M3-1, $R_i$ denotes the product creation rate when the reaction system is at state $\Gamma_i$. $G(\Gamma_j, t | \Gamma_i)$ denotes the propagator, or the conditional probability that the reaction system is at state $\Gamma_j$ at time *t*, given that the system is at $\Gamma_i$ at time 0. $P_{ss}(\Gamma_j)$ is the probability of finding the system at $\Gamma_j$ in the steady state. In the long time limit, $G(\Gamma_j, t | \Gamma_i)$ should approach $P_{ss}(\Gamma_j)$ so that we have $\lim_{t \to \infty} \langle R(t)R(0) \rangle_{ss} = \langle R \rangle^2$. By subtracting $\langle R \rangle^2$ from both sides of Eq. M3-1, we obtain

$$\langle \delta R(t) \delta R(0) \rangle_{ss} = \sum_{i=1}^{l} \sum_{j=1}^{l} R_j \left[ G(\Gamma_j, t | \Gamma_i) - P_{ss}(\Gamma_j) \right] R_i P_{ss}(\Gamma_i). \qquad \text{[M3-2]}$$

For the multi-channel process, the propagator, $G(\Gamma_j, t | \Gamma_i)$, is given by the *j*-th element of the *l*-dimensional column vector $\mathbf{G}(t)$ that satisfies the following time-evolution equation

$$\frac{\partial}{\partial t} \mathbf{G}(t) = \mathbf{T} \cdot \mathbf{G}(t) \qquad \text{[M3-3]}$$

and the following initial condition, $[\mathbf{G}(0)]_j = \delta_{ji}$. In Eq. M3-3, $\mathbf{T}$ represents the transition rate matrix corresponding to the multi-channel reaction, that is,

$$\mathbf{T} = \begin{pmatrix} -k_{21} & k_{12} & 0 & 0 & \cdots \\ k_{21} & -(k_{12}+k_{32}) & k_{23} & 0 & \cdots \\ 0 & k_{32} & -(k_{23}+k_{43}) & k_{34} & \cdots \\ 0 & 0 & k_{43} & -(k_{34}+k_{54}) & \cdots \\ \vdots & \vdots & \vdots & \vdots & \ddots \end{pmatrix}$$

The solution of Eq. M3-3 can be obtained as

$$G(\Gamma_j, t | \Gamma_i) = P_{ss}(\Gamma_j) + \sum_{k=2}^{l} (\mathbf{Q})_{jk} e^{-\lambda_k t} (\mathbf{Q}^{-1})_{ki},$$  [M3-4]

where $\mathbf{Q}$ and $\{\lambda_k\}$ denote the matrix that diagonalizes $\mathbf{T}$, i.e., $\mathbf{Q}^{-1} \cdot \mathbf{T} \cdot \mathbf{Q} = \mathbf{\Lambda}$ with $(\mathbf{\Lambda})_{ij} = \lambda_i \delta_{ij}$, and the non-negative eigenvalues of the transition rate matrix. On the R.H.S. of Eq. M3-3, $P_{ss}(\Gamma_j)$ corresponds to the term with zero eigenvalue, $\lambda_1 = 0$. Substituting Eq. M3-4 into Eq. M3-2, we obtain the following multi-exponential function for the TCF of the product creation rate:

$$\langle \delta R(t) \delta R(0) \rangle_{ss} = \langle \delta R^2 \rangle \sum_{k=2}^{l} c_k \exp(-\lambda_k |t|),$$  [M3-5]

where $\langle \delta R^2 \rangle c_k$ is defined by $\sum_{k=2}^{l} \sum_{i=1}^{l} \sum_{j=1}^{l} R_j (\mathbf{Q})_{jk} (\mathbf{Q}^{-1})_{ki} R_i p_{ss}(\Gamma_i)$. Here, $c_k$ satisfies the normalization condition: $\sum_{k=2}^{l} c_k = 1$.

The power spectrum of the product creation rate can be calculated by taking the Fourier transform of Eq. M3-5:

$$S_R(\omega) = 2\langle \delta R^2 \rangle \sum_{k=2}^{l} c_k \frac{\lambda_k}{\omega^2 + \lambda_k^2}. \qquad \text{[M3-6]}$$

The simplest multi-channel process is the two-channel process in which $R_1 = k$ and $R_2 = 0$. For this simple case, the time correlation function of the reaction rate, given in Eq. M3-5, becomes the simple exponential function,

$$\langle \delta R(t) \delta R(0) \rangle_{ss} = \langle \delta R^2 \rangle \exp(-\lambda t) \qquad \text{[M3-7]}$$

with $\lambda = k_{12} + k_{21}$. Here, $k_{ij}$ denotes the rate of transition from state $\Gamma_i$ to state $\Gamma_j$. For the two-state model, $\langle \delta R^2 \rangle$ is given by $k^2 k_{12} k_{21} / (k_{12} + k_{21})^2$. The power spectrum of the product creation rate for the two-channel process is given by

$$S_R(\omega) = 2\langle \delta R^2 \rangle \frac{\lambda}{\omega^2 + \lambda^2}. \qquad \text{[M3-8]}$$

By substituting Eq. M3-8 into Eq. 3 in the main text, we obtain the power spectrum of product number for the two-channel process:

$$S_z(\omega) = \frac{2\langle R \rangle}{\omega^2 + \gamma^2} \left(1 + \frac{\lambda}{\omega^2 + \lambda^2} F_R \right) \qquad \text{[M3-9]}$$

Eqs. M3-7-M3-9 are used to calculate the theoretical results for the multi-channel process in Fig. 1E-G. The TCF of the product number for the multi-channel process, shown in Fig. 1D, can be calculated by the inverse Fourier transform of Eq. M3-9, whose analytic expression is suppressed here.

Next, we present the derivation of the power spectrum of the product creation rate for the multi-step process shown in Fig.2C. The multi-step product creation process shown Fig. 1C is an example of a renewal process. The analytic expression for the reaction waiting time

distribution of the multi-step product creation process is simple in the Laplace domain and given by(3, 4)

$$\hat{\psi}(s) = \prod_{j=1}^{n} \frac{k_j}{s+k_j}, \quad \quad \text{[M3-10]}$$

where $n$ and $k_j$ denote the number of intermediate reaction steps composing the product creation process and the rate of the $j$-th reaction step, respectively. By substituting Eq. M3-10 into Eq. 5, we can obtain the analytic expression for the power spectrum of the product creation rate for the multi-step reaction process. If we assume that all rates of the internal steps are the same, i.e., $k_l = k$, we can write the rate power spectrum as

$$S_R(\omega) = \frac{2k}{n} \text{Re}\left\{ \frac{1}{[1+i(\omega/k)]^n - 1} \right\}, \quad \quad \text{[M3-11]}$$

where we exploit the fact that $\langle R \rangle = k/n$. Eq. M3-11 then can be rewritten as

$$S_R(\omega) = \frac{2k}{n} \left[ \frac{[1+(\omega/k)^2]^{n/2} \cos n\theta - 1}{[1+(\omega/k)^2]^n - 2[1+(\omega/k)^2]^{n/2} + 1} \right], \quad \quad \text{[M3-12]}$$

where $\theta$ is defined as $\theta = \tan^{-1}(\omega/k)$. Eq. M3-12 and its inverse Fourier transform are used to calculate the power spectrum and the TCF of the product creation rate, respectively, in Fig. 1G and 1F, for the multi-step process. Substituting Eq. M3-12 into Eq. 3 in the main text, we can obtain the power-spectrum of the product number for the multi-step reaction, and the associated TCF of the product number fluctuation by calculating its inverse Fourier transform, which are shown as the blue lines in Fig. 1E and D, respectively.

**Supplementary Method 4 | Simulation method for Fig. 1.**

Here, we provide the algorithms used to simulate the three different creation reactions shown in Fig. 1A-C. We also discuss how to calculate the time-correlation function of the product number, $\langle \delta z(t) \delta z(0) \rangle$, and the power spectrum of the product number, $S_z(\omega)$, from the simulation results.

### A. Simple Poisson birth-death process

When product creation is a Poisson process, its dynamics can be completely characterized by a single rate constant, $R$; the reaction waiting time distribution of the Poisson product creation process is given by $\varphi_r(t) = R e^{-Rt}$. The reaction waiting time, $t_i$, of the $i$-th Poisson product creation event can be generated by $t_i = -R^{-1} \ln u_i$ with $u_i$ being the uniformly distributed random variable between 0 and 1. In Fig. 1, we set the value of $R$ equal to 2 in an arbitrary time unit.

Throughout this work, the product degradation process is assumed to a Poisson process. When each product molecule is generated in the simulation, we also generate the lifetime of each product molecule from the degradation waiting time distribution, $\varphi_d(t) = \gamma e^{-\gamma t}$, where the $\gamma$ is a product's degradation rate constant. That is to say, the lifetime of the $i$-th product molecule is determined from another uniformly distributed random variable, $u_i'$ between 0 and 1 by $\tau_i = -\gamma^{-1} \ln u_i'$. In Fig. 1, we set the value of $\gamma$ equal to 1.

Performing this simulation iteratively, we can generate the number of the time traces of the product number. By taking the average over the time traces of the product number, we obtain the time-dependent mean and variance of the product number. The mean and variance in the product number reach the steady-state values at long times. In our simulation, the steady-state is found to be attained at times longer than $5\gamma^{-1}$.

We calculate the steady-state time autocorrelation function of the product number from the simulation time traces of the product number as follows. We first set the value of $t_0$ as $5\gamma^{-1}$. The mean product number at time $t_0$ is essentially the same as the steady-state value, $\langle z \rangle_{ss}$. For each trajectory, we calculate $\delta z(t_0)$ and $\delta z(t+t_0)$, where $\delta z(t)$ designates $z(t) - \langle z \rangle_{ss}$. By performing the average of $\delta z(t+t_0)\delta z(t_0)$ over the simulation trajectories, we obtain the value of the steady-state TCF, $\langle \delta z(t+t_0)\delta z(t_0) \rangle_{ss}$, of the product number. We confirm that the value of $\langle \delta z(t+t_0)\delta z(t_0) \rangle_{ss}$ is independent of $t_0$, as long as the value of $t_0$ is greater than $5\gamma^{-1}$.

From the simulated time traces of the product number at times longer than $t_0$, the power spectrum of the product number is obtained from a discrete Fourier transform of the deviation of the product number fluctuation, according the definition of the power spectrum, given in Eq. 1 in the main text.

### B. Multi-channel creation process with constant rate decay

In this section, we describe the algorithm used to simulate the multi-channel creation reaction shown in Fig. 1B. As shown in Fig.1B, the reaction state fluctuates with the state-traversing rate, $k_{ij}$, which denotes the state-transition rate constant from the $\Gamma_i$-state to the $\Gamma_j$-state. Each state, $\Gamma_i$, has its own product creation rate constant, $R_i$.

We start the simulation by sampling the initial state using the steady-state state distribution. On sampling the initial state, $\Gamma_i$, we generate the state-transition waiting time by using the transition rate constant from the $\Gamma_i$ state to the adjacent state. If there are two transition directions, we generate the transition waiting time for each direction and choose the shorter one. Once we sample the transition waiting time, we then conduct a simple Poisson product creation reaction simulation by using the state's creation rate constant, $R_i$, until the time reaches the state-transition waiting time. After each product creation process ends, we calculate the product lifetime using the same procedure described in previous section. When the time reaches the previously sampled state-transition time, we change the creation reaction rate constant into the new rate constant, $R_j$, for the new state, $\Gamma_j$. We then repeatedly conduct the sampling of the state-transition waiting time and a simple Poisson product creation simulation using the new rate for the sampled state transition waiting time.

By continually repeating this procedure, we can simulate the multi-channel creation process. Finally, we calculate the time correlation and the power spectrum of the product number fluctuation in the same way as the simple Poisson creation case. For the result of Fig. 1D and E, we choose the simplest multi-channel reaction process, two-state reaction process. When we denote the two-states as the on-state and the off-state, we set the state transition rate from each of the two states as the same as 1/4 in the simulation. The product creation reaction occurs only when our system is at the on-state, and the product creation rate is set to be 4 in our simulation. We set the rate of the product decay process, $\gamma$, as 1.

### C. Multi-step creation process with constant rate decay

To simulate the multi-step reaction process shown in Fig. 1C, we generate the creation reaction time by summing the set of the reaction waiting times of the Poisson reaction steps composing the multi-step reaction process. The other details of the simulation method are the same as the other cases described in the previous sections.

In this simulation, we set the number of the intermediate steps composing a single reaction equal to 20 and, for each step, the catalytic rate in each step to 40, in order to set the mean product creation time to 1/2, the same as the mean product creation time set in the simulation of other reaction schemes shown in Fig. 1. The decay process rate constant, $\gamma$, is set equal to 1, as well.

At long times, the simulated dynamics of the product number statistics reaches the steady-state. However, one can obtain the steady-state product number trajectories from the beginning of the simulation by sampling the first reaction waiting time according to

$$\psi_1(t) = S(t) \bigg/ \int_0^\infty dt' S(t'), \qquad [\text{M4-1}]$$

where $S(t)$ denotes the probability that the multi-step reaction has not yet completed as of time $t$, given that the multi-step reaction started at time 0. Alternatively, we can start our simulation at one of the intermediate reaction steps. The initial sampling probability of the $i$-th intermediate reaction step is given by $\tau_i \big/ \left( \sum_{j=1}^n \tau_j \right)$ with $\tau_i$ being the mean lifetime of the intermediate step, i.e., $\tau_i = k_i^{-1}$.

**Supplementary Method 5 | Analytic expressions for power spectra in Fig. 2.**

In this section, we discuss the power spectra visualized in Fig. 2. According to Eq 3, the protein number power spectrum is related with the power spectrum of the translation rate by

$$S_p(\omega) = S_p^0(\omega) + \frac{S_{R_{TL}}(\omega)}{\omega^2 + \gamma_p^2}, \qquad [\text{M5-1}]$$

where $S_p^0(\omega) = 2\langle R_{TL}\rangle/(\omega^2 + \gamma_p^2)$. Because $R_{TL} = mk_{TL}$ for our model, the above equation can be expressed as

$$S_p(\omega) = S_p^0(\omega) + \frac{k_{TL}^2 S_m(\omega)}{\omega^2 + \gamma_p^2}, \qquad [\text{M5-2}]$$

where $S_m(\omega)$ is the mRNA number power spectrum. Using the Eqs. M5-1 and M5-2, we can find the form of $S_p(\omega)/S_p^0(\omega) - 1$ as

$$\frac{S_p(\omega)}{S_p^0(\omega)} - 1 = \frac{k_{TL} S_m(\omega)}{2\langle m\rangle}, \qquad [\text{M5-3}]$$

where $\langle m\rangle$ is the mean number of mRNA. To obtain the analytic expression of $S_m(\omega)$, we use Eq. 3 again. The result is given by

$$S_m(\omega) = \frac{2\langle R_{TX}\rangle}{\omega^2 + \gamma_m^2} + \frac{S_{R_{TX}}(\omega)}{\omega^2 + \gamma_m^2}. \qquad [\text{M5-4}]$$

Here, the transcription rate is given by $R_{TX} = k_{TX}\xi$, where the value of $\xi$ takes 1 for the active gene state but 0 for the inactive gene state. Using this expression of the translation rate, we obtain $S_{R_{TX}}(\omega)$ as $S_{R_{TX}}(\omega) = k_{TX}^2 S_\xi(\omega)$.

The analytic expression of $S_\xi(\omega)$ can be obtained from the Fourier transform of the time correlation function of the gene state variable $\xi$. According to ref (5), the time correlation function of the gene state variable in our model is given in the Laplace domain by

$$\mathcal{L}[\langle \delta\xi(t)\delta\xi(0)\rangle] \equiv \langle \delta\xi^2\rangle \hat{\phi}_\xi(s)$$
$$= \frac{\tau_{on}\tau_{off}}{(\tau_{on}+\tau_{off})^2}\frac{1}{s} - \frac{1}{s^2(\tau_{on}+\tau_{off})}\frac{[1-\hat{\psi}_{on}(s)][1-\hat{\psi}_{off}(s)]}{1-\hat{\psi}_{on}(s)\hat{\psi}_{off}(s)}, \qquad \text{[M5-5]}$$

where $\tau_{on}$ and $\tau_{off}$ are the first moments of waiting time distributions, $\psi_{on}$ and $\psi_{off}$, respectively. The Fourier transform of this equation is given by

$$S_\xi(\omega) = \langle \delta\xi^2\rangle \lim_{\varepsilon\to 0^+}[\phi_\xi(i\omega+\varepsilon)+\phi_\xi(-i\omega+\varepsilon)] = \langle \delta\xi^2\rangle \tilde{\phi}_\xi(\omega). \qquad \text{[M5-6]}$$

Substituting Eq. M5-5 into Eq. M5-6, we obtain

$$S_{R_{TX}}(\omega) = k_{TX}^2 S_\xi(\omega) = k_{TX}^2 \langle \delta\xi^2\rangle \tilde{\phi}_\xi(\omega) = \frac{\tau_{on}\tau_{off}}{(\tau_{on}+\tau_{off})^2}\pi\delta(\omega) + \frac{k_{TX}^2}{\tau_{on}+\tau_{off}}\frac{\tilde{G}(\omega)}{\omega^2}, \qquad \text{[M5-7]}$$

where $\tilde{G}(\omega)$ is given by $\tilde{G}(\omega) = \lim_{\varepsilon\to 0^+} 2\,\text{Re}[\hat{G}(\varepsilon+i\omega)]$ with

$$\hat{G}(s) \equiv \frac{[1-\hat{\psi}_{on}(s)][1-\hat{\psi}_{off}(s)]}{1-\hat{\psi}_{on}(s)\hat{\psi}_{off}(s)}. \qquad \text{[M5-8]}$$

Substituting the second equality of Eq. M5-7 into Eq. M5-4, we obtain the power spectrum of the mRNA number by

$$S_m(\omega) = \frac{k_{TX}\langle\xi\rangle}{\omega^2+\gamma_m^2}\left[2 + k_{TX}\tilde{\phi}_\xi(\omega)F_\xi\right], \qquad \text{[M5-9]}$$

where $F_\xi = \langle\delta\xi^2\rangle/\langle\xi\rangle$. By substituting Eq. M5-9 into M5-3, we obtain the power spectrum of the protein number as follows:

$$\frac{S_p(\omega)}{S_p^0(\omega)} - 1 = \frac{k_{TL}\gamma_m}{\omega^2 + \gamma_m^2}\left[1 + \frac{1}{2}k_{TX}\tilde{\phi}_\xi(\omega)F_\xi\right]. \qquad \text{[M5-10]}$$

In Eqs. M5-9 and M5-10, $\tilde{\phi}_\xi(\omega)F_\xi$ is given by

$$\tilde{\phi}_\xi(\omega)F_\xi = \frac{\tilde{G}(\omega)}{\tau_{on}\omega^2} \qquad \text{[M5-11]}$$

for nonzero frequency. The definition of $\tilde{G}(\omega)$ can be found below Eq. M5-7.

To obtain the high frequency asymptotic behavior of $S_p(\omega)/S_p^0(\omega) - 1$, we note that the large $s$ limit value of $\hat{G}(s)$ given in Eq. M5-8 is given by $\lim_{s\to\infty}\hat{G}(s) = 1$, because $\lim_{s\to\infty}\hat{\psi}_{on(off)}(s) = 0$. This means that $\tilde{G}(\omega)\left[\equiv 2\lim_{\varepsilon\to 0+}\operatorname{Re}\hat{G}(i\omega+\varepsilon)\right]$ becomes 2 in the high frequency limit. Therefore, $\tilde{\phi}_\xi(\omega)F_\xi$ given in Eq. M5-11 yields the following high frequency asymptotic behavior:

$$\frac{1}{2}\tilde{\phi}_\xi(\omega)F_\xi \xrightarrow{\omega\to\infty} \frac{1}{\tau_{on}\omega^2}. \qquad \text{[M5-12]}$$

With Eq. M5-12 at hand, we obtain the asymptotic behavior of $S_p(\omega)/S_p^0(\omega) - 1$ given in Eq. M5-10 as follows:

$$\frac{S_p(\omega)}{S_p^0(\omega)} - 1 \xrightarrow{\omega\to\infty} \frac{k_{TL}\gamma_m}{\omega^2}. \qquad \text{[M5-13]}$$

**Supplementary Method 6 | Simulation method for Fig. 2.**

Here, we describe the method used to simulate the gene expression network shown in Fig. 2. This simulation consists of the three states: the gene activity fluctuation, the transcription process, and the translation process.

### A. Gene activity fluctuation

The first step in this simulation is to sample the initial state of the gene according the steady-state distribution. The steady-state probability of the active or inactive gene state is given by $p_{on} = k_{on}/(k_{on} + k_{off})$ or $p_{off} = 1 - p_{on}$, where $k_{on}^{-1}$ and $k_{off}^{-1}$ denote the average lifetimes of the inactive and active gene states, respectively, i.e., $k_{on}^{-1} = \int_0^t dt \psi_{off}(t) t \, (\equiv \tau_{off})$ and $k_{off}^{-1} = \int_0^t dt \psi_{off}(t) t \, (\equiv \tau_{on})$. If the uniform random number generated between 0 and 1 is less than $p_{on}$, we choose the active gene state; otherwise, we choose the inactive gene state. Given that the initial gene state is the active or inactive gene state, we generate the lifetime of the active gene state according to the lifetime distribution, $\psi_{on}(t)$ or $\psi_{off}(t)$. To simulate the gene activity fluctuation process, we repeatedly sample the lifetime of the active gene state and the lifetime of the inactive gene state one after the other.

### B. Transcription process with activated gene

The second step we take is to simulate the transcription process. In the model shown in Fig. 2A, the transcriptional rate is given by $R_{TX} = k_{TX}\xi$, where $k_{TX}$ is the transcription rate constant for the gene in the active state. $\xi$ is a stochastic variable whose value is either 1 for the active gene state and 0 for the inactive gene state. To simulate the transcription process, we simulate Poisson transcription events only when the gene state is in the active state. The rest of the simulation method is the same as the simulation method for the multi-channel reaction described in Supplementary method 4B. In this way, we can generate the number of the time traces of the mRNA number. In this simulation, we take the value of $k_{TX}$ as $k_{TX} = 0.51 \text{ min}^{-1}$, which is given in ref. (6).

From the simulation trajectories of the mRNA number, we can calculate various measures of the product number counting statistics, such as the mean, the variance, the time correlation function, and the distribution of the mRNA number, as described in Supplementary Method 4.

### C. Translation process with live mRNA

The last simulation step is the translation process. In the model shown in Fig. 2A, the translation rate is given by $k_{TL}m$. This means that, for each mRNA, proteins are created by the translation rate, $k_{TL}$. To simulate this model, for each mRNA, we simulate the Poisson translation process, or the protein creation process, with the rate $k_{TL}$ during the lifetime of the mRNA, using the same method as that described in Supplementary Method 4A. Because the mRNA lifetime distribution is an exponential function in our model, upon each creation of mRNA, we generate the lifetime of the mRNA by $\tau_i = -\gamma_m^{-1} \ln u_i$, where $u_i$ is a uniformly distributed random variable between 0 and 1. Likewise, upon each generation of protein molecule, the exponentially distributed lifetime of the protein molecule with the mean lifetime $\gamma_p^{-1}$ is generated in the same manner, and this information is then used in generating the simulation trajectories of the protein number. From the simulation trajectories of protein number, we can calculate the power-spectrum and the time-correlation function of the protein number, shown in Fig. 2, using the method described above. In this simulation, we take the values of $k_{TX}$ and $\gamma_p$ as $k_{TX} = 1.695$ min$^{-1}$ and $\gamma_p = 3.18 \times 10^{-2}$ min$^{-1}$, respectively (6).

In the calculation of the power spectrum of the mRNA or the protein number from the simulation trajectories, we perform the discrete Fourier transform with the Kaiser window function (7), which is a commonly used window function for random signal analysis. The Kaiser window function is given by

$$W(n) = \frac{I_0\left(\pi\alpha\sqrt{1-(2n/(N-1)-1)^2}\right)}{I_0(\pi\alpha)}, \qquad \text{[M6-1]}$$

where $I_0(x)$ denotes a modified Bessel function of order 0. In Eq. M6-1, $\alpha$ and $N$ denote the flattening level and the number of data in each time trajectory, respectively. In our work, the values of $\alpha$ and $N$ are set equal to 3 and 50000, respectively.

**Supplementary Method 7 | Power spectrum analysis in Fig. 3.**

In this method, we present the details of the power-spectrum analysis shown in Fig. 3. The gene-expression network model used in the analysis is the same as Fig. 2A. However, instead of assuming the particular gene activation dynamics assumed in Fig. 2B, we use a more general but simpler model of the gene activation process, in which the reaction waiting time distribution of the gene activation, or the waiting time distribution of the inactive gene state, is given by a gamma distribution, $\psi_{off}(t) = t^{a-1}\exp[-t/b]/(\Gamma(a)b^a)$. In this analysis, we assume $\psi_{on}(t)$ as an exponential distribution, i.e. $\psi_{on}(t) = k_a\exp[-k_a t]$, where $k_a$ is the rate of the gene-deactivation process. For this model, the Laplace transform of $\psi_{on}(t)$ and $\psi_{off}(t)$ are given by $\hat{\psi}_{on}(s) = (s+k_a)^{-1}$ and $\hat{\psi}_{off}(s) = (1+sb)^{-a}$, respectively. Substituting these equations into Eq. (M5-7), we can obtain the fully explicit analytic expression of $S_{R_{TX}}(\omega)$ as

$$S_{R_{TX}}(\omega) = \frac{2k_{TX}^2 \tau_{on}^2}{\tau_{on}+\tau_{off}} \frac{1 - \dfrac{\cos[a\tan^{-1}(b\omega)]}{(1+b^2\omega^2)^{a/2}}}{\left(1 - \dfrac{\cos[a\tan^{-1}(b\omega)]}{(1+b^2\omega^2)^{a/2}}\right)^2 + \left(\tau_{on}\omega + \dfrac{\sin[a\tan^{-1}(b\omega)]}{(1+b^2\omega^2)^{a/2}}\right)^2} \quad . \qquad \text{[M7-1]}$$

where $\tau_{off}$ is equal to $ab$. This equation is compared with the data for $S_{R_{TX}}(\omega)$, which can be obtained from the data for the protein number power spectrum, $S_p(\omega)$, by repeated use of Eqs. 6 and 7.

Alternatively, we can directly analyze the protein number power spectrum data. By substituting Eq. M7-1 into Eqs. M5-4, we obtain the power-spectrum, $S_m(\omega)$, of the mRNA number. Then, substituting this equation into Eq. M5-3, we obtain the following expression of $S_p(\omega)/S_p^0(\omega)-1$:

$$\frac{S_p(\omega)}{S_p^0(\omega)} - 1 = \frac{k_{TL}\gamma_m}{\omega^2 + \gamma_m^2}\left[1 + k_{TX}\tau_{on}\frac{1 - \frac{\cos[a\tan^{-1}(b\omega)]}{(1+b^2\omega^2)^{a/2}}}{\left(1 - \frac{\cos[a\tan^{-1}(b\omega)]}{(1+b^2\omega^2)^{a/2}}\right)^2 + \left(\tau_{on}\omega + \frac{\sin[a\tan^{-1}(b\omega)]}{(1+b^2\omega^2)^{a/2}}\right)^2}\right].$$

[M7-2]

This equation can be compared with the data for the protein number power spectrum, $S_p(\omega)$.

In Eqs. M7-1 and M7-2, the adjustable parameters are a, b, and . This is because $\tau_{off}$ is the same as $ab$, and $k_{TX}$ is related to the mean transcription rate $\langle R_{TL} \rangle$ by the following equation:

$$\frac{k_{TX}\tau_{on}}{\tau_{on} + \tau_{off}} = \langle R_{TX} \rangle = \gamma_m \langle m \rangle.$$

[M7-3]

The mean mRNA number can be directly estimated from the mean protein number, $\langle p \rangle$, from $\langle m \rangle = \langle p \rangle / k_{TL}$, where the value of $k_{TL}$ can be, in turn, estimated from the high frequency asymptotic behavior of $S_p(\omega)$ given in Eq. M5-13. Throughout this work, we assume that the values of $\gamma_m$, $\gamma_p$, and $\langle p \rangle$ are provided from the experimental data.

**Supplementary Method 8 | Analytic expressions for power spectrum of the transcription rate in Fig. 4.**

In this method, we present the mathematical details related to the power spectrum analysis given in Fig. 4. The translational part in the gene-expression network model used in Fig. 4 is the same as in Fig. 2A. However, the transcriptional part is here extended to a more general case in which the active gene transcription rate, $k_{TX}$, is treated as a dynamic stochastic variable. The mean value, $\langle k_{TX} \rangle$, of the active gene transcription rate is the same as the value of $k_{TX}$ in Figs. 2 and 3. The normalized time correlation function of $k_{TX}$ is given by an exponentially decaying function of time, i.e., $\phi_{k_{TX}}(t) = \exp(-\lambda t)$. The gene regulation by the promoter is modelled to be the same as Model II in Fig. 3B. The mean-scaled time correlation function of the total transcription rate, $R_{TX} (= k_{TX} \xi)$, can be written as (2)

$$\frac{\langle \delta R_{TX}(t) \delta R_{TX}(0) \rangle}{\langle R_{TX} \rangle^2} = \eta_{k_{TX}}^2 \phi_{k_{TX}}(t) + \eta_{\xi}^2 \phi_{\xi}(t) + \eta_{k_{TX}}^2 \eta_{\xi}^2 \phi_{k_{TX}}(t) \phi_{\xi}(t) \qquad \text{[M8-1]}$$

unless $k_{TX}$ is correlated with $\xi$. The Fourier transform of Eq. M8-1 is given by

$$\frac{S_{R_{TX}}(\omega)}{\langle R_{TX} \rangle^2} = \eta_{k_{TX}}^2 \tilde{\phi}_{k_{TX}}(\omega) + \eta_{\xi}^2 \tilde{\phi}_{\xi}(\omega) + \eta_{k_{TX}}^2 \eta_{\xi}^2 \tilde{\phi}_{k_{TX}}(\omega) * \tilde{\phi}_{\xi}(\omega) \qquad \text{[M8-2]}$$

where $f(\omega) * g(\omega)$ denotes the convolution integral defined by $f(\omega) * g(\omega) \equiv \frac{1}{2\pi} \int_{-\infty}^{\infty} d\omega' f(\omega - \omega') g(\omega')$. Note here that the Fourier transform of $\phi_{k_{TX}}(t) \phi_{\xi}(t)$ can be calculated as

$$\int_{-\infty}^{\infty} dt e^{-i\omega t} \phi_{k_{TX}}(t) \phi_{\xi}(t) = \int_{-\infty}^{\infty} dt e^{-i\omega t} \phi_{k_{TX}}(t) \left[ \frac{1}{2\pi} \int_{-\infty}^{\infty} d\omega' e^{i\omega' t} \tilde{\phi}_{\xi}(\omega') \right]$$

$$= \frac{1}{2\pi} \int_{-\infty}^{\infty} d\omega' \tilde{\phi}_{\xi}(\omega') \int_{-\infty}^{\infty} dt e^{-i(\omega-\omega')t} \phi_{k_{TX}}(t) \qquad [\text{M8-3}]$$

$$= \frac{1}{2\pi} \int_{-\infty}^{\infty} d\omega' \tilde{\phi}_{\xi}(\omega') \tilde{\phi}_{k_{TX}}(\omega - \omega')$$

In terms of the mean-scaled power spectrum, $\tilde{S}_q(\omega) \equiv S_q(\omega) / \langle q \rangle^2$, Eq. M8-2 can be rewritten as

$$\tilde{S}_{R_{TX}}(\omega) = \tilde{S}_{k_{TX}}(\omega) + \tilde{S}_{\xi}(\omega) + \tilde{S}_{k_{TX}}(\omega) * \tilde{S}_{\xi}(\omega) \qquad [\text{M8-4}]$$

which is the same as Eq. 9 in the main text. In the total transcription rate, $R_{TX}(=k_{TX}\xi)$, considered in Supplementary Method 7, the active gene transcription rate, $k_{TX}$, is simply a constant. Therefore, dividing both sides of Eq. M7-1 by $k_{TX}^2$, one can obtain the expression of $S_{\xi}(\omega)$. In case of $S_{k_{TX}}(\omega)$, its expression is simply given by Eq. M3-8 with $R$ being equal to $k_{TX}$. For convenience, the explicit expressions of $\tilde{S}_{\xi}(\omega)$ and $\tilde{S}_{k_{TX}}(\omega)$ are presented below:

$$\tilde{S}_{\xi}(\omega) = 2(\tau_{on} + \tau_{off}) \frac{1 - \frac{\cos[a\tan^{-1}(b\omega)]}{(1+b^2\omega^2)^{a/2}}}{\left(1 - \frac{\cos[a\tan^{-1}(b\omega)]}{(1+b^2\omega^2)^{a/2}}\right)^2 + \left(\tau_{on}\omega + \frac{\sin[a\tan^{-1}(b\omega)]}{(1+b^2\omega^2)^{a/2}}\right)^2} . \qquad [\text{M8-5}]$$

$$\tilde{S}_{k_{TX}}(\omega) = 2\eta_{k_{TX}}^2 \frac{\lambda}{\omega^2 + \lambda^2} . \qquad [\text{M8-6}]$$

**Supplementary Method 9 | Simulation method for Fig. 4.**

Here, we present a detailed description of the simulation method for the gene expression network model shown in Fig. 4A. In this model, the transcriptional rate is given by $R_{TX} = k_{TX}(\Gamma)\xi$, where $k_{TX}(\Gamma)$ is the active gene transcription rate that is dependent on the cell state, $\Gamma$. $\xi$ is a dichotomous stochastic variable whose value is either 1 for the active gene state and 0 for the inactive gene state. To simulate the stochastic fluctuation in $k_{TX}$, the normalized time correlation function of which is given by $\phi_{k_{TX}}(t) = \exp(-\lambda t)$, we adopt the two-state model, where $k_{TX}$ dynamically fluctuates between $k_{TX}^{(1)}[\equiv k_{TX}(\Gamma_1)]$ and $k_{TX}^{(2)}[\equiv k_{TX}(\Gamma_2)]$ with transition rates $k_{12}$ and $k_{21}$. Here, $k_{ij}$ denotes the transition rate from $\Gamma_j$ to $\Gamma_i$. For this model, the mean, noise, and inverse of relaxation time of $k_{TX}$ is given by

$$\langle k_{TX}\rangle = k_{TX}^{(1)}\frac{k_{12}}{k_{12}+k_{21}} + k_{TX}^{(2)}\frac{k_{21}}{k_{12}+k_{21}} \qquad [\text{M9-1}]$$

$$\eta_{k_{TX}}^2 = \frac{k_{12}k_{21}[k_{TX}^{(1)} - k_{TX}^{(2)}]^2}{[k_{TX}^{(1)}k_{12} + k_{TX}^{(2)}k_{21}]^2} \qquad [\text{M9-2}]$$

$$\lambda = k_{12} + k_{21} \qquad [\text{M9-3}]$$

Using the above equations, we can determine the values of the rate parameters that reproduces the given values of $\langle k_{TX}\rangle$, $\eta_{k_{TX}}^2$, and $\lambda$. For each mRNA time trace, initial values of $k_{TX}$ and $\xi$ are sampled with their own steady-state weights and ensuing time traces of $k_{TX}$ and $\xi$ are generated independently of each other. Only when $\xi$ stays at unity over the whole simulation time, mRNAs are produced. The lifetime distributions of the active and inactive gene states are respectively given by $\psi_{on}(t) = k_a e^{-k_a t}$ with $k_a = 0.34 \ \text{min}^{-1}$ and $\psi_{off}(t) = t^{a-1}e^{-t/b}/\Gamma(a)b^a$ with

$a = 6.66$ and $b = 9.73$ min. The ($i$+1)-th mRNA creation time, $t^c_{i+1}$, is sampled from $k_{TX}(t^c_i)e^{-tk_{TX}(t^c_i)}$, where $k_{TX}(t^c_i)$ designates the value of $k_{TX}$ at the $i$-th mRNA creation time, $t^c_i$. When $k_{TX}$ undergoes a transition before a creation event is completed, the incomplete creation event is discarded and the new creation starts at the time of the transition. To calculate the mRNA number power spectrum, we use the same algorithm described in Supplementary Method 6C.

# II. SUPPLEMENTARY NOTES

**Supplementary Note 1| Derivation of Eq. 10.**

Here, we provide a detailed derivation of Eq. 10. The mean-scaled time correlation function of the translation rate, $R_{TL}(=k_{TL}m)$, can be written as (2)

$$\frac{\langle \delta R_{TL}(t) \delta R_{TL}(0) \rangle}{\langle R_{TL} \rangle^2} = \eta_{k_{TL}}^2 \phi_{k_{TL}}(t) + \eta_m^2 \phi_m(t) + \eta_{k_{TL}}^2 \eta_m^2 \phi_{k_{TL}}(t) \phi_m(t) \quad \text{[N1-1]}$$

unless $k_{TL}$ is correlated with $m$. When $\phi_{k_{TL}}(t)$ relaxes much slower than $\phi_m(t)$, Eq. N1-1 can be simplified as

$$\frac{\langle \delta R_{TL}(t) \delta R_{TL}(0) \rangle}{\langle R_{TL} \rangle^2} \cong \eta_{k_{TL}}^2 + (1 + \eta_{k_{TL}}^2) \eta_m^2 \phi_m(t) \quad \text{[N1-2]}$$

The Fourier transform of Eq. N1-2 is given by

$$\begin{aligned}\frac{S_{R_{TL}}(\omega)}{\langle R_{TL} \rangle^2} &= \eta_{k_{TL}}^2 \delta(\omega) + (1 + \eta_{k_{TL}}^2) \eta_m^2 \tilde{\phi}_m(\omega) \\ &= \eta_{k_{TL}}^2 \delta(\omega) + (1 + \eta_{k_{TL}}^2) \frac{S_m(\omega)}{\langle m \rangle^2}\end{aligned} \quad \text{[N1-3]}$$

For nonzero frequency $(\omega > 0)$, Eq. 10 can be obtained by substituting Eq. N1-3 into Eq. M5-1 and noting that $\langle R_{TL} \rangle = \langle k_{TL} \rangle \langle m \rangle$.

# III. SUPPLEMENTARY FIGURES

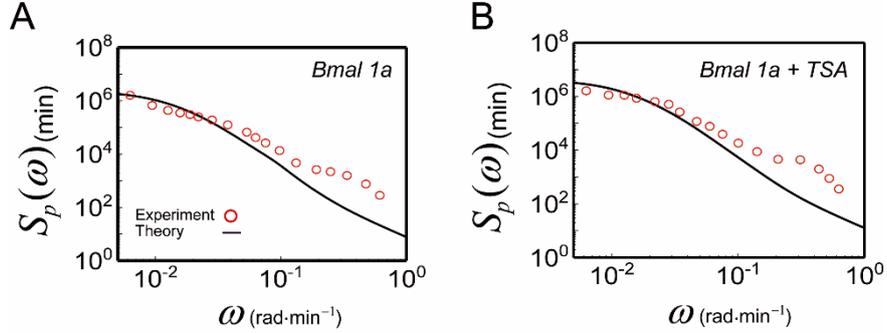

**FIG. S1. Comparison between the model and experiment for the Power spectrum of protein number.** (circles) Power spectrum of the protein number obtained from the protein number time trace data reported by Naef and co-workers (6). In A, we show the power spectrum of the number of luciferase expressed under the control of the *Bmal 1a* promoter in mice fibroblast cells, and in B, we show the power spectrum of the number of luciferase expressed under the same *Bmal 1a* promoter in the same types of cells but in the presence of TSA (trichostatin A) in B. (line) Power spectrum of the protein number calculated from the gene expression network model given in Fig. 2A and B with the rate parameters reported in ref. 5. The protein number in each cell is measured every five minutes for both cases. The observation time ranges of the *Bmal 1a* expression data and the *Bmal 1a* expression with TSA data are 895 minutes and 1180 minutes, respectively. The degradation rate constant of mRNA for each case is given by 0.0102/min. The number of protein number trajectories used in calculation of the protein number power spectrum is 56 in A and 39 in B. Despite the small number of protein number trajectories and/or the strong non-stationarity in the protein number time traces, the protein number power spectrum obtained from the experimental data is in good agreement with the protein number power spectrum calculated from the gene expression network model. A more accurate quantitative analysis of the protein number power spectrum is possible with a greater number of the time traces of the protein number, which can be obtained from a large number of cells in the steady-state.

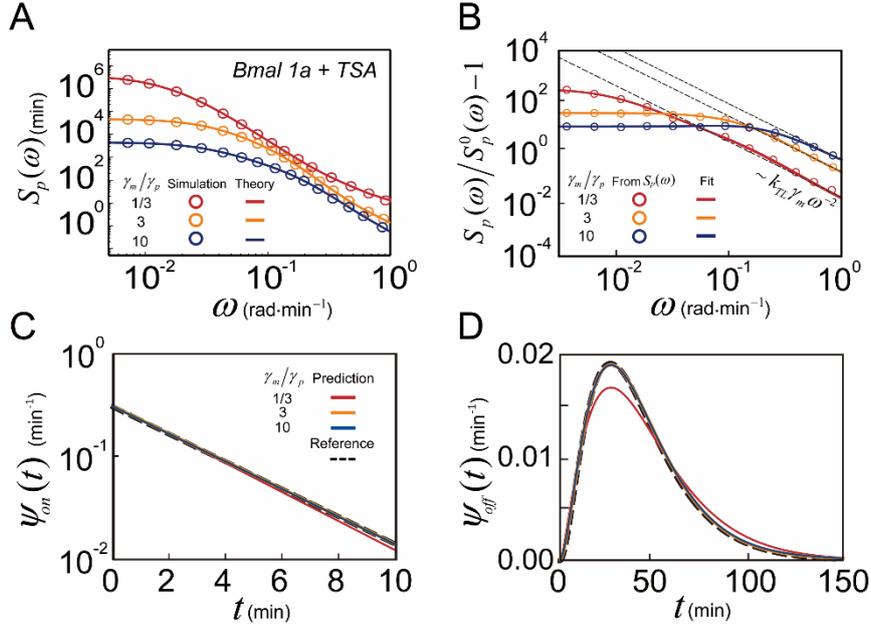

**Fig. S2. Power spectrum analysis of *Bmal 1a* + TSA gene expression.** (A) (circles) Power spectrum calculated from the protein number time traces obtained by the simulation of the gene expression network shown in Fig. 2A and B with use of the reference values of the rate parameters reported in ref (6) for luciferase expression under *Bmail 1a* in mice fibroblast cells. The authors in ref (6) investigated the special case where the lifetime of the luciferase is adjusted to be one-third of its mRNA lifetime. In our simulation, we also conduct the simulation of the gene expression network model for the usual case where the protein lifetime is longer than the mRNA lifetime. It is known that, in the presence of TSA, the number of intermediate reaction steps composing gene activation or the mean lifetime of the inactive gene state decreases. To analyze this shortening, Naef and co-workers propose that the number of gene activation steps is three, instead of seven. By using the reference value for the rate parameters other than the protein decay rate, we conduct simulations for the gene expression network shown in Fig. 2A and B to obtain the time traces of mRNA and protein numbers. (line) Prediction of our theory for the same gene expression network model. (B) Quantitative analysis

of the simulation data for $S_p(\omega)/S_p^0(\omega)-1$. $S_p^0(\omega)$ is defined as $S_p^0(\omega)=2\langle R_{TL}\rangle/(\omega^2+\gamma_p^2)$. Given the value of $\gamma_p$, one can estimate the value of the mean translation rate, $\langle R_{TL}\rangle$, from the mean protein number, $\langle p\rangle$, by $\langle R_{TL}\rangle=\gamma_p\langle p\rangle$. In the quantitative analysis of the data, we have used Eq. M7-2, which is obtained for the gene expression network model in Fig. 2A with the Poisson gene deactivation process and the non-Poisson gene activation process with the reaction waiting time distribution given by a gamma distribution. The values of the extracted parameter are given by $a=2.86$, $b=15.43$, and $\tau_{on}=4.37\,\text{min}$ for $\gamma_m/\gamma_p=1/3$, $a=2.78$, $b=15.13$, and $\tau_{on}=3.21\,\text{min}$ for $\gamma_m/\gamma_p=10$. (C, D) Protein number power spectrum results. Note that $\psi_{on}(t)$ extracted from this analysis is approximately the same as $\psi_{on}(t)$ shown in Fig. 3E, but $\psi_{off}(t)$ extracted from this analysis has a smaller mean than $\psi_{off}(t)$ extracted from the analysis shown in Fig. 3F.

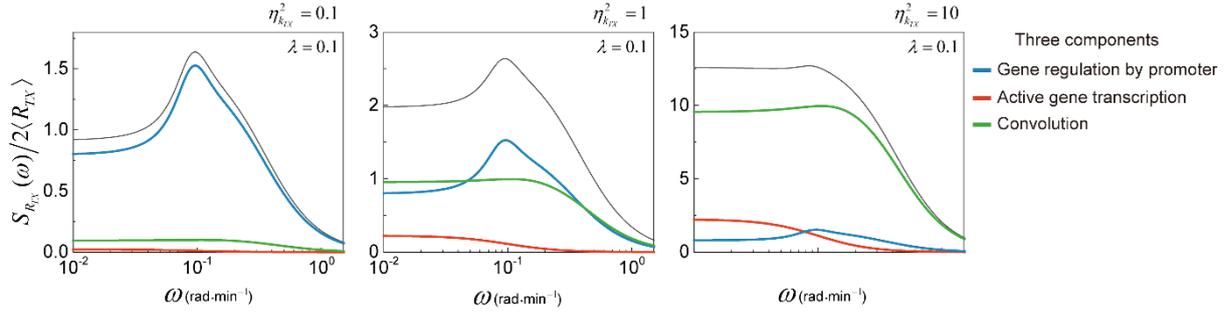

**Fig. S3. Three components in the power spectrum of the transcription rate.** (black line) The value of $S_{R_{TX}}(\omega)/2\langle R_{TX}\rangle$ or $S_m(\omega)/S_m^0(\omega)-1$ with $S_m^0(\omega)=2\langle m\rangle\gamma_m/(\omega^2+\gamma_m^2)$. The value of $\lambda$ is here set to be 0.1 min$^{-1}$. (colored lines) Three components of the power-spectrum: $\langle R_{TX}\rangle\tilde{S}_\xi(\omega)/2$ originating from the gene regulating dynamics of promoter (blue), $\langle R_{TX}\rangle\tilde{S}_{k_{TX}}(\omega)/2$ originating from the active gene transcription dynamics (red), and their convolution $\langle R_{TX}\rangle\tilde{S}_\xi(\omega)*\tilde{S}_{k_{TX}}(\omega)/2$ (green). See Eqs. 9 or M8-4. Unlike the case shown in Fig. 4C, the peak originating from the promoter fluctuation diminishes with $\eta_{k_{TX}}^2$ because a fast fluctuation in $k_{TX}$ with a rate $\lambda$ comparable to or larger than the promoter fluctuation frequency, $\omega_{peak}=2\pi/(\tau_{on}+\tau_{off})\cong 0.1$ rad·s$^{-1}$, effectively filters out the contribution of the promoter fluctuation in $\langle R_{TX}\rangle\tilde{S}_\xi(\omega)*\tilde{S}_{k_{TX}}(\omega)/2$.

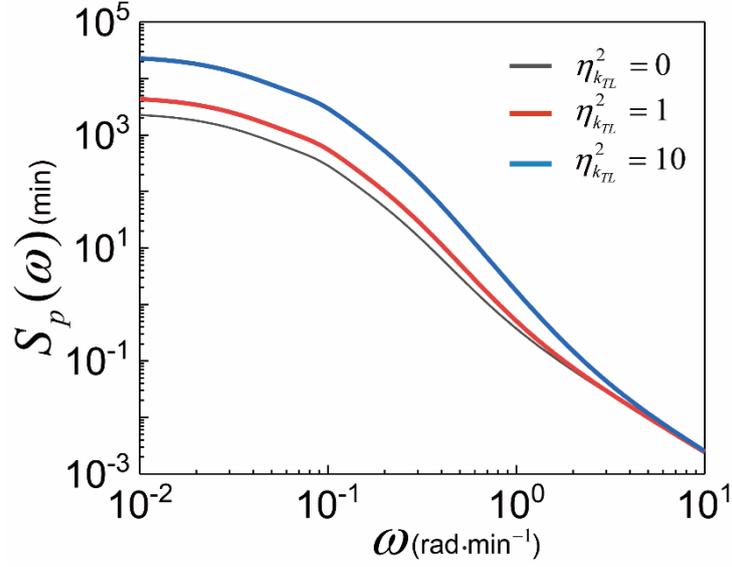

**Fig. S4. Effect of the cell-to-cell heterogeneity in translation rate coefficient, $k_{TL}$, on the protein number power spectrum, $S_p(\omega)$.** The relevant expression of $S_p(\omega)$ is reproduced here: $S_p(\omega) = S_p^0(\omega)\left[1 + \langle k_{TL}\rangle(1+\eta_{k_{TL}}^2)S_m(\omega)/2\langle m\rangle\right]$, which is equivalent to Eq. 10. $S_p^0(\omega)$ and $S_m(\omega)$ denote the protein number power spectrum without any fluctuation in translation rate, $R_{TL}(=k_{TL}m)$ and the mRNA number power spectrum, respectively. As shown in the equation given above, an increase of the relative variance, $\eta_{k_{TL}}^2$, of $k_{TL}$ amplifies the contribution of the mRNA number power spectrum to the protein number power spectrum. As a result, the protein number power spectrum increases with $\eta_{k_{TL}}^2$. Here, the protein number power spectrum with $\eta_{k_{TL}}^2 = 0$ is the same as that given in Fig. 2C when the value of $\gamma_m/\gamma_p$ is equal to 10. However, the protein number power spectrum is not essentially affected by $\eta_{k_{TL}}^2$ over the high frequency range in which $S_p^0(\omega)$ dominates the other term, $\langle k_{TL}\rangle(1+\eta_{k_{TL}}^2)S_p^0(\omega)S_m(\omega)/2\langle m\rangle$.

# IV. SUPPLEMENTARY REFERENCES